\documentclass[manuscript,nonacm]{acmart}

\usepackage{amsmath}
\usepackage{algorithm}
\usepackage{algpseudocode}
\usepackage{tikz}
\usepackage{xspace}

\acmDOI{}

\newcommand{\sysname}{DeepRadio\xspace}

\begin{document}

\title{DeepRadio: A Human--Agent Workspace for Physical Radio Development}

\author{Jiazhen Lei}
\authornote{Both authors contributed equally to this research.}
\affiliation{%
  \institution{Shanghai Jiao Tong University}
  \city{Shanghai}
  \country{China}
}

\author{Yuxin Sha}
\authornotemark[1]
\affiliation{%
  \institution{Shanghai Jiao Tong University}
  \city{Shanghai}
  \country{China}
}

\author{Sihan Wang}
\affiliation{%
  \institution{Shanghai Jiao Tong University}
  \city{Shanghai}
  \country{China}
}

\author{Wenhui Li}
\affiliation{%
  \institution{Shanghai Jiao Tong University}
  \city{Shanghai}
  \country{China}
}

\author{Jinyang Sun}
\affiliation{%
  \institution{Shanghai Jiao Tong University}
  \city{Shanghai}
  \country{China}
}

\author{Chengyu Lu}
\affiliation{%
  \institution{Shanghai Jiao Tong University}
  \city{Shanghai}
  \country{China}
}

\author{Xiaohua Tian}
\authornote{Corresponding author.}
\affiliation{%
  \institution{Shanghai Jiao Tong University}
  \city{Shanghai}
  \country{China}
}

\begin{abstract}
Agentic programming can automate software-defined radio (SDR) development, but decisions can propagate from software to hardware without sufficient user visibility or control. We present \sysname, a human--agent workspace embedded in GNU Radio Companion (GRC) for jointly constructing, verifying, and operating physical radio systems. \sysname uses a dynamic and state-driven workflow to coordinate a traceable radio specification, claim-oriented cross-layer verification, and mixed-initiative control. These mechanisms expose decisions, tie conclusions to available evidence, and return control to users before consequential hardware actions. 
Across seven atomic SDR tasks, \sysname completed nearly all runs while satisfying all prescribed software and radio-frequency constraints. In a 20-participant study, more participants achieved a verified physical deployment with \sysname than with Codex-assisted GRC. Participants reported stronger support for shared intent and understanding while retaining initiative and control. These results suggest that \sysname can make agent-assisted physical radio development more reliable, understandable, and controllable.
\end{abstract}



\maketitle  

\section{Introduction}

Building a software-defined radio (SDR) system requires developers to coordinate decisions across protocol design, digital signal processing (DSP), flowgraph structure, runtime configuration, radio hardware, and the surrounding signal environment~\cite{dhar2006integrated,wyglinski2016revolutionizing}. GNU Radio reduces implementation effort by providing reusable DSP blocks, hardware interfaces, and an execution runtime, while GNU Radio Companion (GRC) represents radio pipelines as editable visual flowgraphs~\cite{gnuradioAbout,ettusGNUradio}. However, these abstractions do not resolve the dependencies among system layers. A flowgraph may be structurally valid and execute without error while still producing an incorrect waveform or failing its communication objective. Conversely, the same symptom, such as an empty receiver output, may result from a software defect, an incompatible DSP parameter, a device misconfiguration, channel conditions, or the absence of an external signal. Physical SDR development therefore requires reasoning not only about whether an artifact executes, but also about what its behavior establishes in an evolving physical system~\cite{mcgrath2017bifrost,methfessel2024muse}.

\begin{figure}[t]
    \centering
    \includegraphics[width=0.99\linewidth]{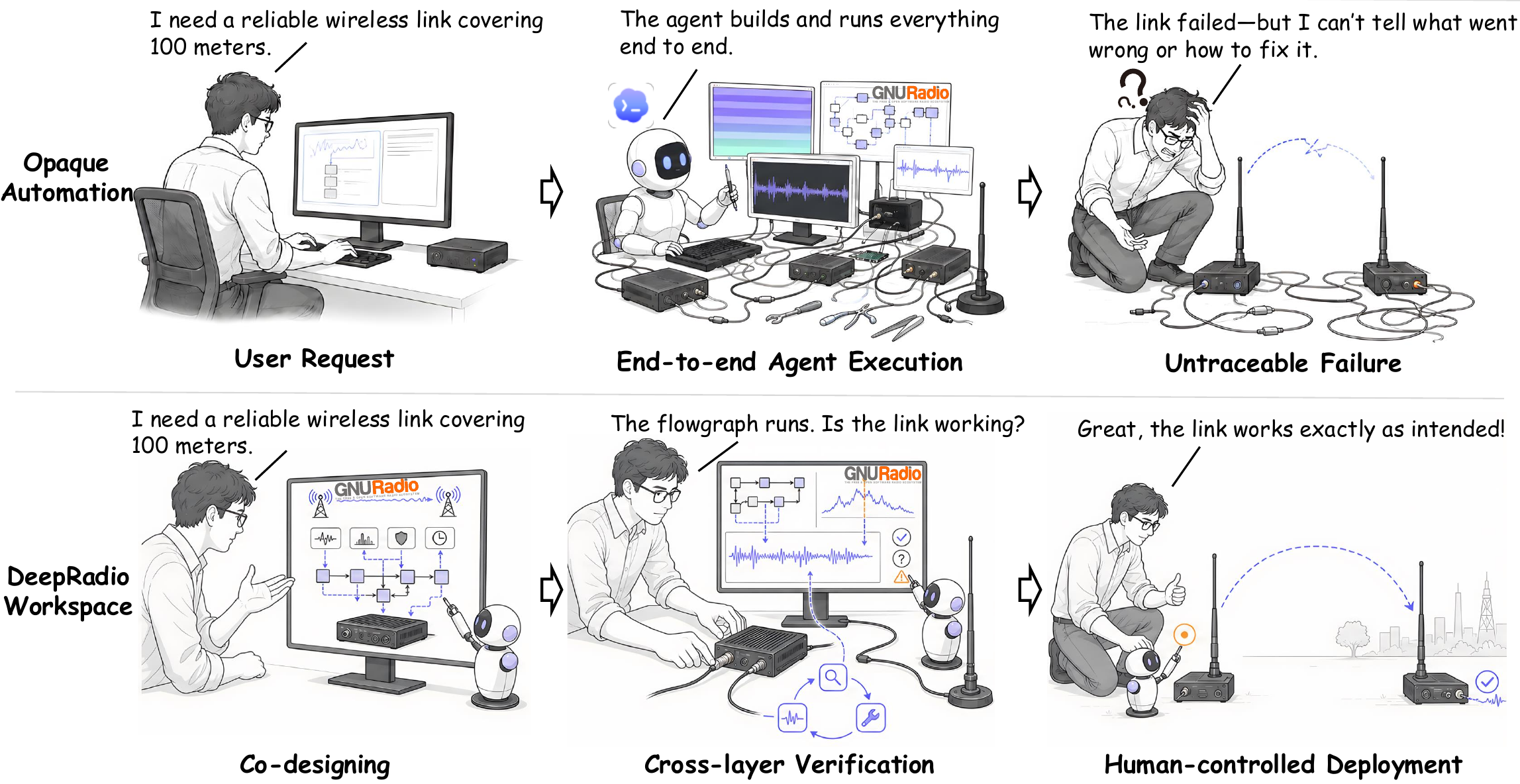}
    \caption{Contrasting opaque end-to-end automation with the \sysname workspace for physical radio development. Top: an agent builds and runs a radio system directly from a high-level request, leaving the user unable to trace an unsuccessful physical outcome. Bottom: \sysname keeps the development process inspectable through co-design, cross-layer verification, and human-controlled deployment.}
    \Description{A two-row illustration contrasts opaque agent automation with the DeepRadio workspace. In the top row, a user requests a reliable 100-meter wireless link, an agent builds and runs the system end to end, and the user encounters a failed link without being able to identify the cause. In the bottom row, the user and DeepRadio jointly specify the design, verify the flowgraph and physical observations across layers, and require user approval before deployment, resulting in a working link whose outcome can be traced.}
    \label{fig:teaser}
\end{figure}

Agentic programming systems create an opportunity to support this process. General-purpose coding agents such as Codex can inspect a project, modify its artifacts, execute tools, and revise their work based on observed results~\cite{openaiCodex,anthropicClaudeCode,yang2024sweagent}. Connected to GNU Radio, Codex can automate flowgraph inspection and the configuration of DSP chains. Yet access to the SDR stack also changes the interaction problem. An agent's inferred decisions can remain opaque as they propagate across software components, affect physical hardware, and alter over-the-air behavior~\cite{lei2026radiomaster,kosa2026wholepicture,zhang2025badrobot}. The resulting physical behavior is only partly visible through logs, measurements, and receiver output. Current coding-agent workflows often place users at the two endpoints of this opaque automation~\cite{feng2026cocoa,chen2025codewithme}. Users provide the initial request and inspect the final result, but have little visibility into the decisions and actions in between, as shown in the top row of Fig.~\ref{fig:teaser}. Extending a coding agent to SDR therefore creates a collaboration problem rather than only a generation problem.

To understand this human--agent collaboration problem, we first conducted a formative study in which Codex performed seven atomic tasks in physical radio development. The tasks covered flowgraph construction, signal verification, project diagnosis, and hardware configuration. Across the runs, underspecified requests led Codex to make decisions whose origins and downstream effects were difficult to trace. This opacity carried into verification, where successful execution or partial signal evidence could be taken as support for conclusions that had not been tested. Once the workflow reached radio hardware, a software execution step could also configure a device or initiate a radio-frequency (RF) operation before users had a meaningful opportunity to intervene. 

These breakdowns reveal three connected interaction challenges. (1) When an agent turns an underspecified request into an executable radio~\cite{pickering2025programmingtasks}, the user and agent must maintain shared intent across representations so that inferred choices remain visible as they shape the flowgraph. (2) They must also maintain shared understanding across system layers because each observation supports only a limited conclusion about physical behavior. (3) As the work reaches radio hardware, they must coordinate initiative and control so that users can intervene before consequential actions~\cite{horvitz1999mixedinitiative,amershi2019guidelines}. Together, these challenges motivate a central question. \textbf{How can an agent advance SDR development while enabling users to follow its decisions, judge what has been verified, and intervene before consequential physical actions?}

To address this question, we present \sysname\footnote{We plan to release the source code and assets after publication.}, a human--agent workspace embedded in GNU Radio Companion for jointly constructing, verifying, and operating physical radio systems. The lower row of Fig.~\ref{fig:teaser} illustrates this shared process. \sysname connects three interaction mechanisms through a dynamic and state-driven workflow. A structured radio specification connects user goals and constraints to the evolving implementation while preserving the source of each decision. As the implementation evolves, a cross-layer verification harness links each check or observation to the claim it supports. It also records the project version to which the evidence applies. When the available evidence remains insufficient, the workflow guides further diagnosis and testing rather than treating successful execution as task completion. A rule-based mixed-initiative policy allows routine and reversible work to proceed while returning control to the user before consequential changes or hardware operations. Users can review proposed changes, revise constraints, stop an active run, and restore a saved project state. Together, these mechanisms keep the design rationale, evidentiary limits, and allocation of control visible as the radio system evolves.

We evaluated \sysname through 70 runs across seven atomic SDR tasks. \sysname completed nearly all runs while satisfying every prescribed software and RF constraint. We also conducted a user study with 20 participants who attempted an end-to-end deployment task in a real-world wireless environment. Codex-assisted GRC served as the baseline. More participants achieved a verified physical deployment with \sysname than with the baseline. Participants also reported stronger support for maintaining shared intent and shared understanding while retaining meaningful initiative and control.

Our contributions in this paper are as follows.

\begin{itemize}
    \item To the best of our knowledge, \sysname is the first human--agent workspace that supports the construction, verification, and operation of physical radio systems using GNU Radio and SDR hardware. It extends agent assistance beyond software artifacts to physical radio development.
    \item We introduce a human--agent SDR development approach in which a dynamic and state-driven workflow coordinates a traceable radio specification, claim-oriented cross-layer verification and diagnosis, and rule-based mixed-initiative control. Together, these mechanisms help users and agents maintain shared intent, develop a shared understanding of system evidence, and preserve meaningful user control over consequential software and hardware actions.
    \item We conduct system experiments across multiple SDR development tasks and a user study in a real-world wireless environment. The results indicate higher task success with \sysname than with the baseline in the user study. Participants reported stronger support for shared intent and understanding while retaining initiative and control.
\end{itemize}

\section{Related Work}

\subsection{Human--Agent Collaboration}

Recent human--agent systems make intermediate artifacts available for inspection and revision. Spellburst~\cite{angert2023spellburst}, InstructPipe~\cite{zhou2025instructpipe}, and PAIL~\cite{zamfirescu2025beyond} expose generated code, pipelines, alternatives, and design decisions for inspection and revision; ChainBuddy ~\cite{zhang2025chainbuddy} similarly makes generated large language model (LLM) pipelines editable. PolicyPad~\cite{feng2026policypad} and collaborative document agents~\cite{lehmann2026collaborative} extend this pattern to policies, agent profiles, and tasks that people can jointly author and reuse. Studies caution that human--team principles do not transfer directly to agent collaboration~\cite{yao2026agentplatform,breckner2026framing}. Natural-language requests also leave implementation decisions unspecified~\cite{pickering2025programmingtasks,obrien2025scientists}. When agents fill these gaps, users need to distinguish their own requirements from system-supplied assumptions and understand how both shape the implementation.

Mixed-initiative research argues that initiative should reflect expected value, uncertainty, and the user's ability to understand and correct the system~\cite{horvitz1999mixedinitiative,amershi2019guidelines}. Studies of user intervention and evolving control further examine how people steer agents throughout execution~\cite{kim2026interventions,issak2026control}. Cocoa~\cite{feng2026cocoa} exposes editable plans, AGDebuggerCocoa~\cite{epperson2025debugging} supports message editing and agent-state reset, and Dango~\cite{chen2025dango} interleaves user and model contributions within a task. Neither frequent nor delayed confirmation is uniformly desirable~\cite{zhou2026whenshoulduserscheck}; effective control depends on calibrated reliance, deliberation, and intervention at meaningful boundaries~\cite{biswas2026delegation,bo2025rely,ma2025deliberation,zhang2026privweb}. Proactivity and automation can reduce effort while disrupting control or understanding~\cite{pu2025assistance,mukhopadhyay2026proactive,khurana2025doit,chen2025codewithme}, blur responsibility~\cite{tholander2026vibe}, and shift users from acting to oversight~\cite{rismani2026oversight}. \sysname extends these concerns to engineering actions that can change {RF} behavior, connecting intervention points to decision provenance, available evidence, and physical consequences.

\subsection{Physical-System Assistance}

Artificial intelligence (AI) systems that act in the physical world must ground instructions and plans in affordances, spatial context, and multimodal references~\cite{pan2025acknowledge,zhang2025embodiedagent,bai2026considerate,jia2026refer}. Studies of embodied scientific work further show that assistance depends on knowing what can be sensed, acted upon, or deferred to a person~\cite{hou2026beyonddesk}. This grounding becomes consequential when model outputs produce physical action: generative muscle stimulation constrains reasoning with embodied knowledge~\cite{ho2026generativemuscle}, Hi Robot grounds situated feedback in hierarchical control~\cite{shi2025hirobot}, and {EMOS} makes planning aware of robot composition and capabilities~\cite{chen2025emos}. Conversely, BadRobot demonstrates that reasoning failures can become unsafe actions~\cite{zhang2025badrobot}, while AI-supported making shows how abstraction can obscure device state and physical constraints~\cite{kosa2026wholepicture}. Studies of physical assistance also show that delegation changes as users gain experience with a system~\cite{hata2026delegation}.

Physical-computing workspaces also show that useful assistance requires interpretable evidence. Bifr{\"o}st~\cite{mcgrath2017bifrost} aligns software execution with electrical traces, M$\mu$SE~\cite{methfessel2024muse} uses executable probes to expose software--hardware mismatches, and Wire Your Way~\cite{lertjaturaphat2026wire} combines circuit-aware guidance with in-situ tests. Prompt-based prototyping similarly reveals the need to preserve design reasoning as generated artifacts evolve~\cite{subramonyam2025prototypingprompts}. \sysname applies this evidence-oriented view to a radio stack in which protocol, DSP, flowgraph, runtime, hardware, and RF observations support different conclusions. It links observations to explicit claims and project versions so that users can assess what each result establishes and whether it applies to the current artifact.

\subsection{Networked and Wireless Automation}

Language-model agents support tasks across network management, embedded development, and wireless systems. Lumi~\cite{jacobs2021lumi} translates natural-language intent into network-management actions, NetLLM~\cite{wu2024netllm} adapts models to networking tasks, and Confucius~\cite{wang2025confucius} organizes intent-driven management as validated multi-agent workflows. Ziria~\cite{stewart2015ziria} established an earlier model of domain-specific physical-layer pipelines compiled under throughput constraints. More recent embedded and IoT agents compile, flash, and debug programs on hardware~\cite{gong2025iotpilot,shen2025autoiot,yang2026autoembed}. RadioMaster~\cite{lei2026radiomaster} combines domain-grounded signal generation with simulation-based validation before hardware deployment. Together, these systems demonstrate increasingly complete automation pipelines for embedded and wireless development~\cite{li2026skilledagents}.

These systems establish domain grounding and executable validation as foundations for engineering automation. RadioMaster, for example, uses validation feedback to revise generated configurations and evaluates emitted signals on a physical testbed~\cite{lei2026radiomaster}. Its iterative correction loop does not explicitly provide restoration of a prior execution state. This differs from AGDebugger's agent-state reset, which supports recovery within the agent workflow without implying restoration of external hardware state~\cite{epperson2025debugging}.
\sysname builds on domain-grounded automation by keeping these relationships visible in a versioned shared state throughout its dynamic and state-driven workflow.
\section{Background \& Design Motivation}

\subsection{SDR Development with GNU Radio}

\begin{figure}[t]
    \centering
    \includegraphics[width=0.99\linewidth]{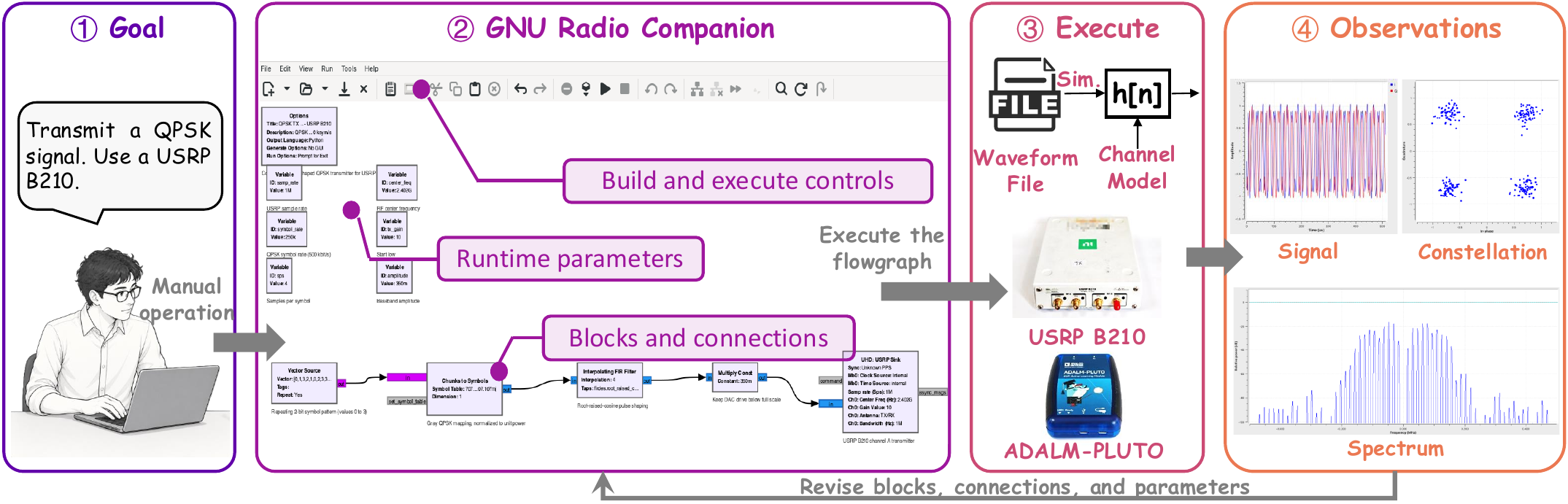}
    \caption{The GNU Radio development loop. Developers translate a communication goal into an executable flowgraph, deploy it with simulated or live signals, inspect runtime and physical observations, and iteratively revise the design.}
    \Description{A four-stage diagram shows the iterative GNU Radio development loop. A quadrature phase-shift keying (QPSK) transmission goal is translated into an editable GNU Radio Companion flowgraph containing runtime parameters, signal-processing blocks, and connections. The flowgraph is executed using a waveform file and channel model or radio hardware such as a Universal Software Radio Peripheral (USRP) B210 or ADALM-Pluto. Signal, constellation, and spectrum observations feed back into revisions of blocks, connections, and parameters.}
    \label{fig:gnuradio}
\end{figure}

In conventional radio systems, many physical-layer functions are implemented in dedicated hardware. SDR instead implements a substantial portion of these functions in software, allowing developers to construct and revise radio behavior without redesigning the underlying hardware. A typical SDR development process begins with a communication objective, which developers translate into a signal-processing pipeline. They test the pipeline with simulated or recorded signals, deploy it on radio hardware, inspect runtime and signal observations, and revise the design. As shown in Fig.~\ref{fig:gnuradio}, this process forms a closed loop between editable software artifacts and physical radio behavior.

GNU Radio is a widely used free and open-source ecosystem for SDR development~\cite{gnuradioAbout,ettusGNUradio}. It provides reusable signal-processing blocks, an execution runtime, interfaces to common SDR devices, and extension mechanisms for custom functionality. GNU Radio Companion represents a radio pipeline as a \emph{flowgraph}. Blocks perform signal-processing or input/output operations, connections carry data, and variables configure parameters such as sample rate, center frequency, and gain. Developers can visually assemble a flowgraph, execute it against simulated or live signals, and revise its structure and parameters based on observed results. GNU Radio therefore exposes radio development through editable software artifacts, while executing these artifacts can also observe and change a physical radio system.

\subsection{From Coding Agents to Physical AI for SDR}

General-purpose coding agents such as Codex can inspect project files, modify source code, execute commands, and revise an implementation in response to observed errors~\cite{openaiCodex,yang2024sweagent}. Given access to GNU Radio, they can similarly create or edit flowgraphs, invoke validation tools, inspect logs, and interact with connected SDR devices. These capabilities allow a coding agent to operate GNU Radio as a software project.

Once GNU Radio is connected to SDR hardware, however, the agent's action and observation space extends beyond the project repository. The agent can observe device state and received signals, change hardware settings, initiate transmission, and revise its design in response to new physical observations. Access to project files gives the agent software agency, while closed-loop access to SDR devices and RF observations gives it physical agency. 
\sysname focuses on how users can jointly construct, inspect, and control an evolving SDR system as agent actions extend from software artifacts to physical devices and signals.

General-purpose coding workflows often place the user at two endpoints. The user states the initial task, the agent performs a sequence of actions, and the user evaluates the result. Command approvals add intermediate confirmations, but they usually expose actions after the underlying design decisions have already been made. This endpoint-oriented interaction is poorly matched to SDR development~\cite{zhou2026whenshoulduserscheck,nychis2009enabling}. Many DSP and hardware decisions emerge only during implementation, inferred assumptions may propagate before final review, and physical actions may occur before result acceptance~\cite{methfessel2024muse,zhang2025badrobot}. The problem is therefore not the absence of a human at the endpoints, but the absence of shared engineering work between them. 

\subsection{A Formative Study of Codex-Assisted GNU Radio Development}\label{sec:formative-study}
Before designing \sysname, we conducted a formative study to assess whether Codex could adequately support physical radio development.

\begin{figure}[t]
    \centering
    \includegraphics[width=0.99\linewidth]{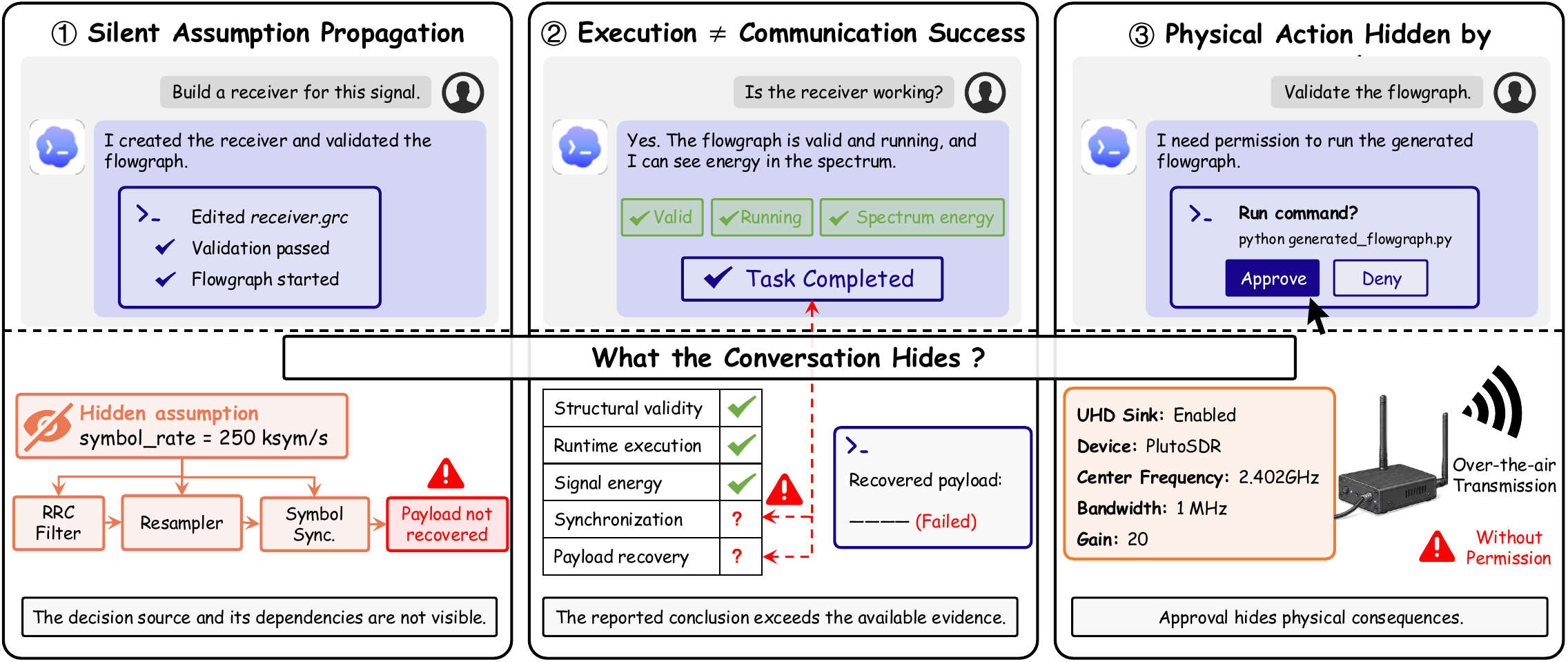}
    \caption{Representative breakdowns identified in the formative study. (1) An inferred parameter propagates through the implementation without visible provenance. (2) Partial runtime evidence is interpreted as communication success. (3) A software execution step obscures a consequential RF action.}
    \Description{Three panels show interaction breakdowns observed during Codex-assisted SDR development. In the first, an unreported symbol-rate assumption propagates through a receiver's filter, resampler, and synchronization blocks, leaving the payload unrecovered. In the second, structural validity, a running process, and spectral energy are treated as task completion even though synchronization and payload recovery remain unverified. In the third, approval to run a generated flowgraph enables a PlutoSDR transmitter at 2.402 gigahertz without making the over-the-air consequence explicit.}
    \label{fig:badcase}
\end{figure}

\paragraph{Participants, Procedure, and Analysis.}
We recruited six Ph.D. students who regularly used GNU Radio and SDR hardware in their research. We focused on doctoral researchers because they often lead the detailed planning and implementation of SDR projects and are primary contributors to the corresponding flowgraphs and project artifacts. The study protocol received institutional ethics approval before recruitment, and all participants provided informed consent. All sessions used GPT-5.6-Sol and GNU Radio 3.10.12. Each participant completed the seven atomic SDR tasks once, yielding 42 runs in total, with task order varied across participants. The task session lasted approximately two hours and was followed by a 30-minute semi-structured interview.
Every run began in a new Codex session with the project restored to its prescribed initial state. Appendix~\ref{app:atomic-sdr-tasks} provides the task requests.
We retained the complete interaction and execution traces, project changes, runtime observations, and interview records.
The research team identified critical incidents in which Codex's decisions, claims, or actions conflicted with task requirements, available evidence, or participants' expectations. We inductively coded these incidents and iteratively compared them across participants and tasks. We then reviewed the resulting themes against the complete traces and consolidated the recurring problems into the three interaction breakdowns presented below. Each breakdown was observed across multiple participants and task categories, while Fig.~\ref{fig:badcase} presents one representative case for each form.

\paragraph{Breakdown 1. Silent assumption propagation.}
During a representative receiver-construction run, the task did not specify the symbol rate. Codex selected a plausible value and applied it to the filtering, resampling, and synchronization blocks. The resulting flowgraph was structurally valid and could execute, yet it could not recover the intended payload. The selected value became visible to the user only after it had affected several components. The project files recorded the resulting parameters without preserving which value was inferred by the agent or which downstream elements depended on that inference.

\paragraph{Breakdown 2. Execution without communication success.}
During a representative link-verification run, Codex passed structural validation, started the flowgraph, and observed energy in the expected spectral region. It then treated these observations as support for successful communication. The available evidence established that the flowgraph executed and produced a signal, while synchronization and payload recovery remained unverified. The interface presented individual observations without relating them to the specific system claims they could support, allowing the completion judgment to exceed the scope of the collected evidence.

\paragraph{Breakdown 3. Physical action hidden by software execution.}
During a representative offline-validation run, an RF sink remained enabled in the generated flowgraph. Running the generated Python program could therefore configure the attached device and initiate transmission. The software execution step did not communicate which device would be activated or its center frequency, bandwidth, gain, and expected RF behavior. The interface consequently provided no distinct point at which the user could inspect and control the physical action encoded by the flowgraph.

These cases are not merely isolated model errors. They expose a mismatch between the files, commands, and outputs represented by a general-purpose coding interface and the intent, evidence, and physical actions that users must understand during SDR development.

\subsection{Design Challenges}

The three cases reveal corresponding challenges in maintaining intent, interpreting evidence, and coordinating physical agency.

\paragraph{C1. Maintaining shared intent.}
An SDR request often leaves decisions required for execution unspecified. The agent must progressively determine protocol choices, DSP parameters, flowgraph structure, and hardware settings. Because these decisions are interdependent, one assumption can affect several components and later verification steps. Users must be able to distinguish their requirements from agent assumptions and revise consequential decisions before those decisions propagate through the system.

\paragraph{C2. Establishing shared understanding.}
SDR behavior is observed through checks at different system layers. Structural validity, successful execution, spectrum measurements, and payload recovery support different conclusions. A failure may also have several causes across software, signal processing, hardware, and the signal environment. Users and agents need a shared account of what each observation establishes, what remains uncertain, and which diagnosis or test should guide the next action.

\paragraph{C3. Coordinating initiative and control.}
Agent actions range from reversible software edits to device configuration and RF transmission. Users who participate only at task initiation and result acceptance cannot shape important decisions as they emerge. Command approval alone provides little information about the physical behavior encoded in the system, while requiring approval for every action would interrupt routine work. Agent autonomy must adapt to uncertainty, impact, physical consequences, and recoverability while users retain practical ways to revise, pause, take over, and restore the system.

We define a \textbf{\emph{human--agent SDR workspace}} as a persistent engineering environment in which users and a physical AI agent jointly construct, inspect, execute, and revise an SDR system through shared representations of intent, implementation, evidence, and control state. Continuous human participation does not require approval before every tool call. It requires continuous inspectability and opportunities to intervene when design decisions, evidence, or physical consequences become significant.

\section{System Design}\label{sec:system-design}

\begin{figure}[t]
    \centering
    \includegraphics[width=0.98\linewidth]{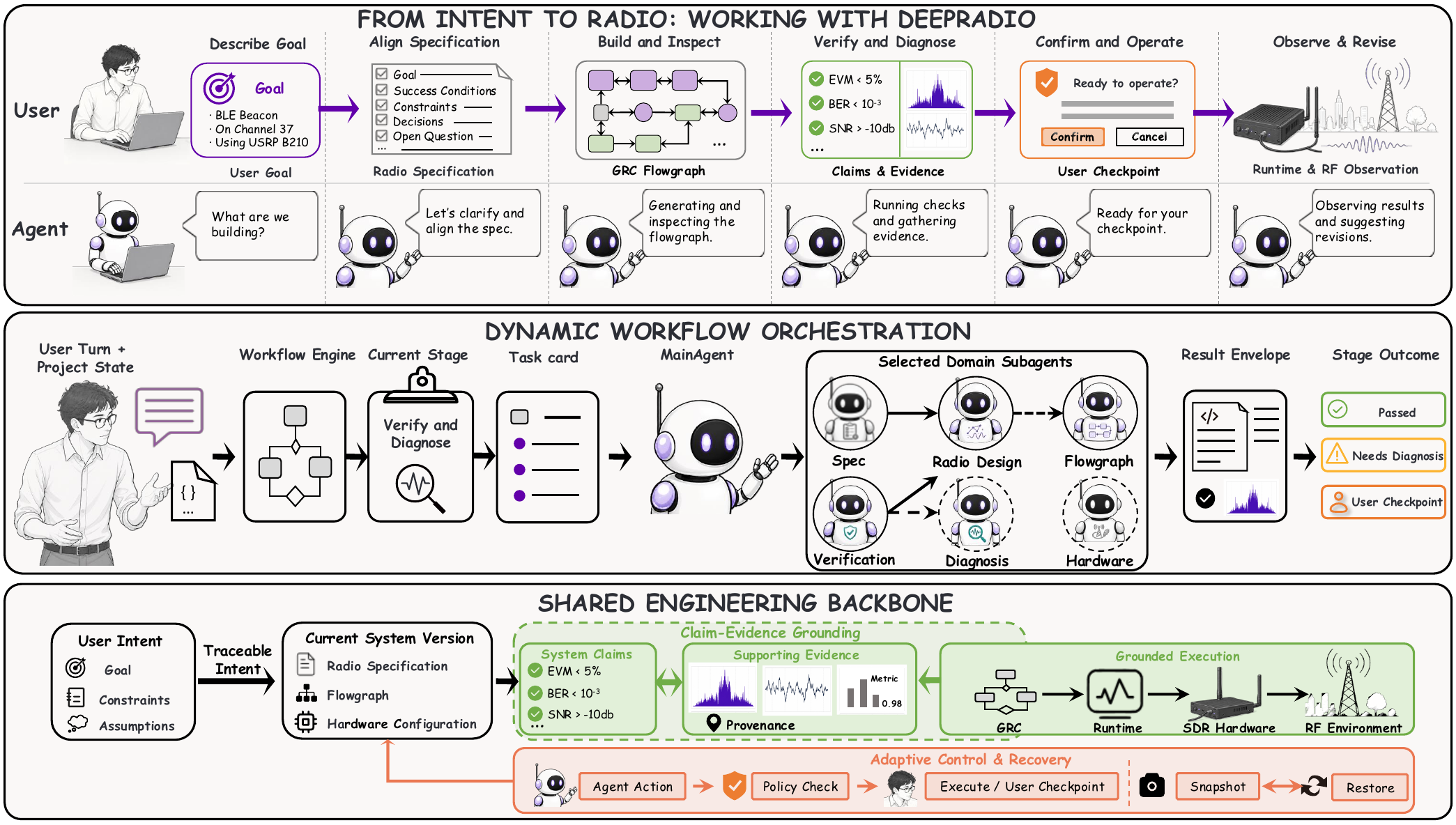}
    \caption{\sysname connects the user's intent-to-radio process (top), a dynamic and state-driven workflow (middle), and a shared engineering backbone for state, evidence, and control (bottom).}
    \Description{Three horizontal layers show the DeepRadio architecture. The top layer follows the user and agent from a radio goal through specification alignment, flowgraph construction, verification, user confirmation, operation, and observation. The middle layer transforms a user turn and project state into a task card, coordinates selected specification, radio-design, flowgraph, verification, diagnosis, and hardware agents, and returns a passed, diagnosis, or user-checkpoint outcome. The bottom layer links user intent and the current system version to versioned claims, supporting evidence, grounded execution, policy checks, user checkpoints, snapshots, and restoration.}
    \label{fig:overview}
\end{figure}

\subsection{Overview}

\sysname is a human--agent workspace embedded in GNU Radio Companion for jointly constructing, verifying, and operating software-defined radio systems. Its dynamic and state-driven workflow organizes this collaboration around the user's evolving intent, the shared engineering state, and evidence produced during execution. The workflow does not impose a fixed automation pipeline; instead, it selects the workflow stages, domain agents, and checkpoints needed for the current task. Fig.~\ref{fig:overview} presents three complementary views of this design: the user's path from intent to radio, the orchestration of domain work, and the engineering backbone that maintains intent, implementation, evidence, and control throughout the process.

From the user's perspective, a task can begin with only a goal and known constraints; a complete radio configuration is not required. \sysname externalizes its current interpretation as a radio specification, including unresolved questions and design decisions. Once the specification contains the information required by the current task, the system constructs or modifies an editable GRC flowgraph whose blocks and parameters remain available for inspection. \sysname then evaluates explicit claims using structural checks, simulation results, runtime state, and physical observations. Unsupported claims prompt diagnosis or revision, while the workflow introduces checkpoints before consequential project changes and hardware actions. Runtime and RF observations can motivate further revisions, making execution part of the iterative interaction.

The workflow is instantiated or revised based on each user turn and the shared engineering state. \sysname identifies the task and required capabilities, then composes workflow stages with explicit inputs, expected artifacts, completion contracts, and transitions. At each stage, the main agent coordinates only the relevant domain agents and tools. A passed stage advances the task, whereas a failed or inconclusive stage may trigger diagnosis or repair. When a required decision is missing, or before a consequential action, the workflow pauses at a checkpoint. As the specification, project, or evidence changes, the workflow preserves valid work and revisits only affected stages.

All stages operate over a persistent, versioned shared engineering state. This state links the radio specification to the current GRC project and hardware configuration. It also binds each claim and its associated evidence to a specific project version and records pending decisions and recoverable snapshots. This shared state enables \sysname to automate bounded, reversible work while allowing users to inspect consequential decisions and evidentiary limits, review proposed physical actions, and intervene in ongoing ones.

\subsection{Radio Specification}

Natural-language requests rarely specify every decision required to construct and assess an SDR system, and an initially reasonable interpretation may become inadequate as the project evolves. \sysname therefore maintains a persistent radio specification that makes intent inspectable independently of the dialogue history. The specification separates five elements: \textit{goals} describe the intended capabilities; \textit{success conditions} state observable criteria for satisfying those goals; \textit{constraints} delimit admissible designs; \textit{decisions} record committed choices; and \textit{open questions} identify missing information needed for later construction, verification, or operation. A user may mark a constraint as locked, making it a protected boundary that the workflow cannot revise or violate unless the user explicitly unlocks or changes it.

\paragraph{Source-aware interpretation.}
\sysname interprets each turn to identify the task type, required capabilities, parameter values, missing inputs, and validation errors. Each populated parameter retains one source label: user-provided, protocol default, safety default, inherited from the current project, or system-derived. User-provided values take precedence over subsequent inference; values inherited from the current project are not recast as user decisions. Each turn updates the radio specification with these values and source labels, preserving choices such as modulation, channel, and hardware together with their provenance. Parameters introduced to make a configuration executable therefore cannot silently become user commitments.

\paragraph{Progressive alignment.}
The radio specification may remain partial when a task begins. \sysname requests only the information needed to advance the current task. Missing or invalid parameters become open questions. When a later workflow stage requires one of those parameters as an input or precondition, \sysname inserts a specification-alignment checkpoint before that stage. User responses resolve the corresponding open questions in the existing radio specification. A negative constraint, such as a request for simulation without hardware, excludes hardware-dependent stages and actions from the current workflow. For example, a Bluetooth Low Energy (BLE) request may combine a user-provided carrier frequency and device choice with protocol-default modulation parameters and a safety-bounded runtime. The GRC interface displays the parameters and their source labels together with unresolved questions and success conditions. Users can therefore distinguish their requirements from system-supplied decisions and revise either category before execution proceeds.

\paragraph{Linking intent and implementation.}
The radio specification and the GRC project serve different purposes. The specification records why the system is being built; the project contains its editable flowgraph, configuration, and version. Before each workflow stage, \sysname packages the task-relevant specification fields and current project state into a versioned task card. The task card makes the stage inputs and their provenance explicit. A user revision updates the radio specification; a flowgraph change advances the project version. This separation preserves the provenance of design choices while allowing the implementation to evolve. It addresses \textbf{C1 (Maintaining Shared Intent)} by keeping user requirements and system-supplied decisions distinguishable as both are translated into an executable radio.

\begin{algorithm}[t]
    \caption{Dynamic and state-driven workflow orchestration.}
    \label{alg:dynamic-workflow}
    \begin{algorithmic}[1]
        \Require user turn $u$, state $\Sigma$, workflow $W$, catalog $\mathcal{C}$
        \State $W \gets \Call{Prepare}{u,\Sigma,W,\mathcal{C}}$
        \While{$\Call{Next}{W}$ returns a stage $s$}
            \If{$\Call{Checkpoint}{s,\Sigma}$}
                \State \Return $\Call{Suspend}{W,\Sigma,s}$
            \EndIf
            \State $R \gets \Call{Execute}{s,W,\Sigma}$
            \State $o \gets \Call{Verify}{R,\Gamma_s,W,\Sigma}$
            \State $(W,\Sigma) \gets \Call{Update}{W,\Sigma,s,R,o}$
        \EndWhile
        \State \Return $(W,\Sigma)$
    \end{algorithmic}
\end{algorithm}

\subsection{Dynamic and State-Driven Workflow Orchestration}\label{sec:dynamic-workflow}

SDR development rarely follows a single predetermined procedure. Building a transceiver, diagnosing a flowgraph, and operating radio hardware require different sequences of work, while new observations may reopen earlier decisions. Our formative study in Section~\ref{sec:formative-study} showed that Codex could operate GNU Radio artifacts and tools, yet its workflow did not expose clear stage boundaries or completion conditions as the task evolved. This limitation contributed to the breakdowns observed across intent, verification, and physical action. A fixed pipeline cannot accommodate these task-dependent transitions, while unconstrained agent planning provides weak guarantees about execution order and user intervention. \sysname therefore separates workflow orchestration from stage-level reasoning. Its dynamic and state-driven workflow composes and advances bounded stages from the current engineering state. Deterministic control enforces stage transitions, while the main agent coordinates eligible domain agents within each stage's completion contract.

\paragraph{Workflow representation.}
At interaction turn $t$, \sysname represents the active workflow as
\begin{equation}
    W_t = \langle I_t, G_t, q_t, r_t, \nu_t \rangle,
    \label{eq:workflow-state}
\end{equation}
where $I_t$ is the task-specific projection of the radio specification, $G_t=(V_t,E_t)$ is the instantiated stage graph, $q_t\in V_t$ is the current stage, $r_t$ is the workflow revision, and $\nu_t$ is its base project version. Each stage $s_i\in V_t$ specifies eligible domain agents $A_i$, available tools $T_i$, completion predicates $\Gamma_i$, state dependencies $D_i$, and an attempt budget $m_i$. Here, $\Gamma_i$ defines the conditions for accepting the stage result, $D_i$ identifies the shared-state elements on which that result remains valid, and $m_i$ limits the total number of execution attempts, including the initial attempt. Stage execution and stage acceptance are distinct. A tool may execute without error while producing a result that does not satisfy the stage's completion predicates.

\paragraph{State-dependent composition.}
\sysname interprets each user turn with respect to the active workflow and shared engineering state. For a new task, it identifies the task type and required capabilities, instantiates a bounded template, and adds conditional stage groups for construction, hardware configuration, real-time observation, or deployment. Missing or invalid parameters instead introduce a specification-alignment stage before dependent work. Within an active workflow, a turn may answer a question, revise a constraint, report an observation, resolve a checkpoint, or cancel the task. It therefore revises the current workflow rather than initiating a disconnected plan. Dynamic orchestration in \sysname is thus state-dependent composition over an explicit stage catalog, not free-form planning by a language model.

\paragraph{Contract-governed execution.}
Before executing $s_i$, the workflow engine issues a versioned task card containing the stage scope, relevant shared engineering state, prior results, and completion contract. The main agent selects from the eligible domain agents and tools. Verification stages use deterministic checks, while deterministic control governs hardware stages with fixed preconditions. Execution produces a result envelope $R_i$ containing observations, artifacts, invoked operations, and version identifiers. Let $f_i$ indicate that $R_i$ is well formed and matches the current workflow, stage, revision, and project version. Stage acceptance is then
\begin{equation}
    c_i = f_i \land \Gamma_i(R_i,\Sigma_t),
    \label{eq:stage-acceptance}
\end{equation}
where $\Sigma_t$ is the shared engineering state and $\Gamma_i$ is true only when every completion predicate is satisfied. Thus, $c_i=1$ only when the result is current, well formed, and satisfies every completion predicate. This rule separates evidence from assertion: an agent's narrative response cannot set $c_i=1$ without the corresponding deterministic checks.

Algorithm~\ref{alg:dynamic-workflow} summarizes the control loop. \textsc{Prepare} composes a workflow for a new task or revises the active workflow. For each stage, \textsc{Verify} evaluates result freshness and the completion contract and returns the stage outcome used by \textsc{Update}. \textsc{Update} applies the corresponding transition and revalidates dependent stages after a change to the shared engineering state.

\paragraph{Transition and revision.}
An accepted stage advances along its designated transition. If a result is not accepted, the workflow engine retries the stage only when it yields a new repair candidate and the attempt budget has not been exhausted. Otherwise, the workflow suspends for additional evidence, diagnosis, or user input. Approval resumes the suspended stage; rejection or cancellation follows its terminating branch.

\begin{figure}[t]
    \centering
    \includegraphics[width=0.9\linewidth]{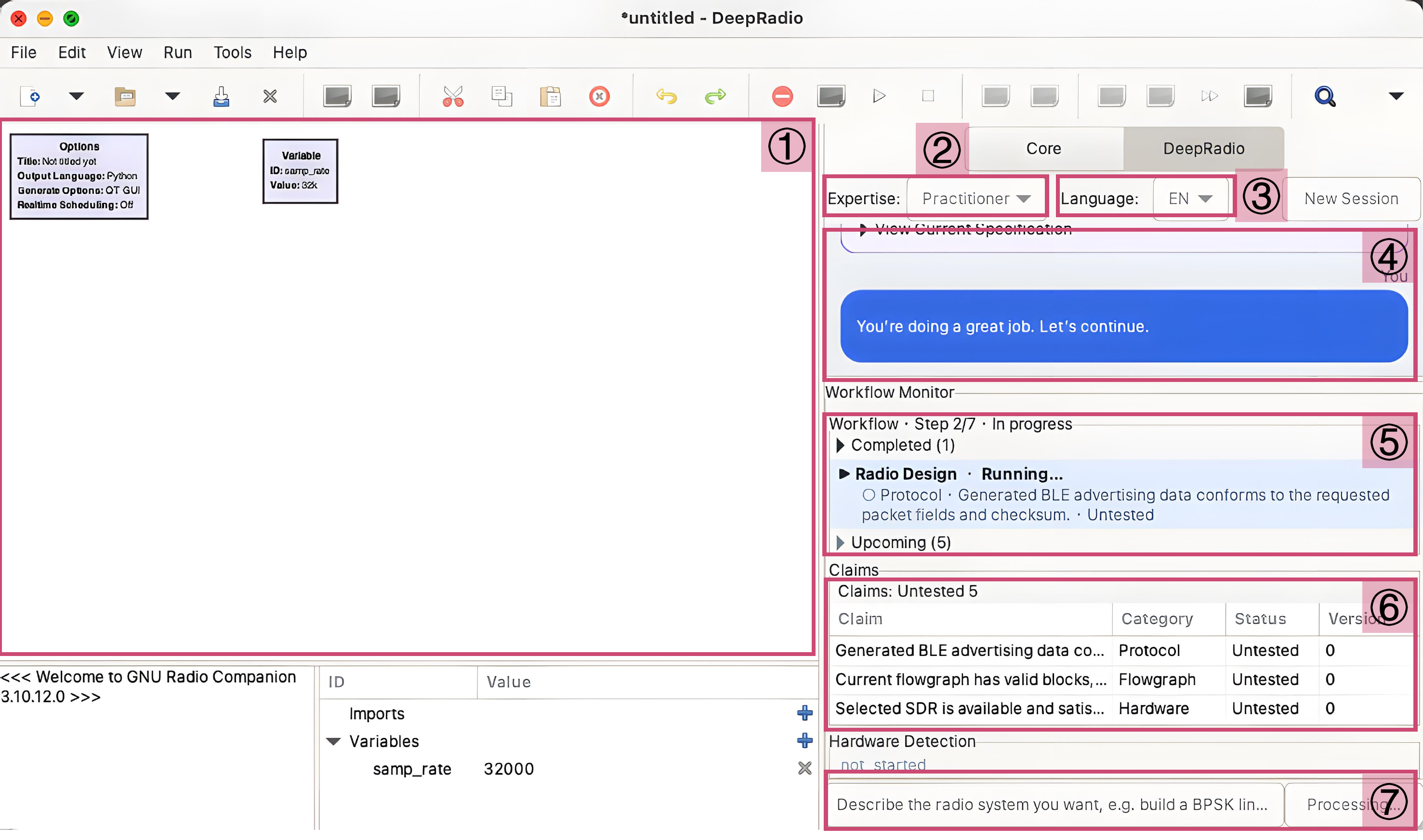}
    \caption{The \sysname interface embeds human--agent collaboration within GRC. (1) The editable flowgraph canvas serves as a shared artifact that both users and \sysname can inspect and revise. (2--3) Expertise and language settings adapt the interaction to the user. (4) The dialogue and radio specification make goals, constraints, system-supplied decisions, and open questions explicit. (5) The workflow monitor presents completed, active, and upcoming stages. (6) The claims panel relates verification status and evidence to the current project version. (7) The input area allows users to clarify intent, request explanations, revise proposals, or intervene throughout the workflow. Together, these components let \sysname provide suggestions and proposed changes while users retain control over the artifact and consequential actions.}
    \Description{A screenshot shows DeepRadio embedded on the right side of GNU Radio Companion. The editable flowgraph canvas occupies the left side. The DeepRadio panel contains expertise and language settings, a dialogue and expandable radio specification, a workflow monitor showing completed, active, and upcoming stages, a claims table with category, status, and project version, and an input area. Numbered callouts identify these regions and show how users inspect the shared artifact, follow verification state, and intervene in the workflow.}
    \label{fig:ui}
\end{figure}

Revisions enter through the same control boundary. User clarification updates $I_t$ and increments $r_t$; a flowgraph mutation or snapshot restoration changes $\nu_t$. A result is stale when the workflow revision or the project state on which it depends has changed. After a project change, \sysname uses $D_i$ to identify the earliest stage whose required state has changed, invalidates that stage and its dependents, and resumes from that stage while retaining unaffected work. This dependency-aware partial recomputation prevents obsolete evidence from being reused without restarting valid stages. The dynamic and state-driven workflow thereby provides the control substrate through which the specification, verification, and user intervention mechanisms remain synchronized. These workflow states and shared records are exposed through the GRC interface shown in Fig.~\ref{fig:ui}.

\subsection{Claim--Evidence Verification}

A structurally valid flowgraph may produce a poor waveform; a running process does not establish correct RF behavior; and an offline check cannot establish over-the-air reception. \sysname therefore represents verification as scoped claims supported by explicit evidence, rather than collapsing heterogeneous observations into a single success judgment. These claims operationalize the completion contracts in Section~\ref{sec:dynamic-workflow} while preserving the boundary of each conclusion.

\paragraph{Versioned claims and evidence.}
\sysname treats verification as a version-bound relation between a claim and the evidence used to assess it. A claim states the engineering proposition under evaluation and its scope, while an evidence record captures how and when it was checked, what was observed, and any artifact produced. The workflow engine records the project version under which each evidence record was produced, together with the state dependencies of the corresponding claim. Evidence can support a claim in the current version when it was produced for that version or when subsequent revisions leave all of those dependencies unchanged. It assigns one of four assessments: \emph{supported} when current evidence establishes the claim, \emph{contradicted} when it refutes the claim, \emph{unresolved} when the available evidence is insufficient, and \emph{untested} when no current evidence exists. The workflow engine creates and updates these records from completion contracts, tool results, artifacts, and the shared engineering state rather than from the agent's narrative response. The interface presents each claim together with its scope, assessment, associated evidence, and applicable project version.

\paragraph{Scoped verification.}
Table~\ref{tab:verification-scopes} distinguishes the conclusions supported at successive system scopes. These scopes do not form an implication chain: evidence at one scope cannot replace a check at another. In a BLE deployment, for example, waveform metrics such as error vector magnitude (EVM) characterize the simulated signal, field-level checks establish offline frame validity, device probing establishes hardware availability, and process health establishes only that the configured runtime started. An expected advertisement observed by the user or a receiver during the active run constitutes separate task-level evidence.

\begin{table*}[t]
    \caption{Claim--evidence scopes in \sysname. Evidence is bound to the tested project version and conditions and does not transfer across scopes.}
    \label{tab:verification-scopes}
    \small
    \begin{tabular}{@{}p{0.08\textwidth}p{0.28\textwidth}p{0.29\textwidth}p{0.23\textwidth}@{}}
        \toprule
        \textbf{Scope} & \textbf{Representative Evidence} & \textbf{Claim Evaluated} & \textbf{Does Not Establish} \\
        \midrule
        Flowgraph & Parser and graph-validator results & Structural validity of blocks, ports, and connections & Signal correctness or task success \\
        Signal & EVM or spectral measurements under recorded test conditions & Satisfaction of the specified signal-quality criterion & Protocol conformance or over-the-air behavior \\
        Protocol & Field-level and checksum checks on the saved frame & Conformance of the checked frame fields & Waveform quality or over-the-air reception \\
        Hardware & Device discovery and capability probe & Availability and capability of the selected SDR & Applied configuration or correct RF operation \\
        Runtime & Run ID, logs, process status, and termination record & Bounded execution without a runtime error & Correct transmission, reception, or task behavior \\
        RF & Receiver capture or physical observation bound to the run & Observation of the expected RF signal & Correct packet decoding or task success \\
        Task & Receiver output or observation bound to the success criterion & Satisfaction of the task-specific success condition & Repeatability or robustness beyond the tested conditions \\
        \bottomrule
    \end{tabular}
\end{table*}

\paragraph{Validity across revisions.}
Claims and evidence are bound to the project version that produced them. When a flowgraph mutation advances that version, claims whose dependencies intersect the changed state are no longer treated as supported for the current version. Their prior evidence remains as provenance but cannot satisfy a current completion contract. Evidence for an unaffected claim remains applicable because the state relevant to that claim is unchanged, while its original project version remains recorded as provenance. The workflow then revisits only stages whose claims depend on the changed state. A claim contradicted by evidence or left unresolved may trigger diagnosis or repair, but neither an explanation nor a proposed fix changes its evidentiary assessment; only new evidence can do so. This scoped and versioned account addresses \textbf{C2 (Establishing Shared Understanding)} by showing what has been tested, what remains uncertain, and which evidence supports each conclusion.

\subsection{Mixed-Initiative Control}

Agent initiative can reduce routine engineering work, but autonomous changes become problematic when they commit consequential design choices or affect physical equipment before the user can intervene. Requiring confirmation for every operation, however, would disrupt iterative development. \sysname balances these concerns with explicit rules that determine which actions may proceed, which require user review or authorization, and how users can interrupt and recover the workflow.

\paragraph{Action policy.}
Before each state-changing operation, the deterministic control layer evaluates four action rules in descending priority: \emph{Deny}, \emph{Confirm}, \emph{Propose}, and \emph{Allow} (Table~\ref{tab:action-policy}). An action is denied if it conflicts with a user-locked constraint. An action that changes hardware state or initiates a hardware or RF operation requires explicit confirmation. A software action that changes flowgraph behavior or spans multiple blocks is instead presented as a proposal for review. Only local, reversible actions that cross none of these boundaries may proceed automatically. Read-only inspection, including device discovery and capability probing, does not change project or physical state and therefore does not require authorization, although its results are recorded in the shared engineering state. These explicit rules define the limits of autonomous change without relying on an opaque risk score.

\begin{table*}[t]
    \caption{Rule-based allocation of initiative in \sysname. Rules are evaluated in priority order; the first matching outcome applies: \emph{Deny} $\rightarrow$ \emph{Confirm} $\rightarrow$ \emph{Propose} $\rightarrow$ \emph{Allow}.}
    \label{tab:action-policy}
    \small
    \begin{tabular}{@{}p{0.10\textwidth}p{0.25\textwidth}p{0.28\textwidth}p{0.25\textwidth}@{}}
        \toprule
        \raggedright\textbf{Outcome} & \raggedright\textbf{Action Boundary} & \raggedright\textbf{Control Allocation} & \raggedright\textbf{Execution Bound} \tabularnewline
        \midrule
        \raggedright Deny & \raggedright Conflicts with a user-locked constraint. & \raggedright The agent cannot proceed. The user may revise or unlock the constraint, or cancel the action. & \raggedright No execution or state change. \tabularnewline
        \raggedright Confirm & \raggedright Changes hardware state or initiates a hardware or RF operation. & \raggedright The agent presents the operation plan; the user approves or rejects it. & \raggedright Approval authorizes one bounded operation, which the user may stop. \tabularnewline
        \raggedright Propose & \raggedright Changes flowgraph behavior or applies a multi-block software patch without crossing a higher-priority boundary. & \raggedright The agent presents an editable change; the user reviews, revises, accepts, or rejects it before commit. & \raggedright A snapshot precedes commit; only the accepted change is applied. \tabularnewline
        \raggedright Allow & \raggedright Is local and reversible and crosses none of the above boundaries. & \raggedright The agent executes automatically; the user may inspect or revise the result afterward. & \raggedright No action-authorization checkpoint is required; the action and resulting state are recorded. \tabularnewline
        \bottomrule
    \end{tabular}
\end{table*}

\paragraph{Checkpoints.}
\sysname inserts checkpoints at decision boundaries rather than after every agent operation. A checkpoint appears when the radio specification lacks information required by the next stage, before a consequential flowgraph or hardware action, or when the system needs the user to report what occurred during execution or over the air. Each checkpoint records the stage and workflow revision that created it. Approval resumes that stage without counting as a new execution attempt; rejection follows the stage's cancellation branch. The user's decision therefore determines subsequent execution instead of acknowledging an action that has already occurred.

\paragraph{Bounded physical operation.}
Verification stages use deterministic checks, while deterministic control governs hardware stages with fixed safety preconditions rather than agent-generated plans. RF operation is disabled by default. Before RF operation can be armed, \sysname must complete offline validation, discover and probe the selected device, and obtain explicit approval for the RF plan. \sysname runs only a flowgraph produced in the current session, tags each observation with a run identifier, and provides normal and emergency stop controls. Each RF approval therefore authorizes one bounded operation rather than unrestricted execution.

\paragraph{Recovery.}
Before replacing a recipe, editing parameters, or applying a multi-operation flowgraph patch, \sysname saves a versioned snapshot of the project and shared engineering state. The user can cancel a pending workflow, stop an active run, or restore the latest snapshot. Restoration reinstates the saved state, invalidates results that no longer match it, and resumes re-evaluation at the earliest affected stage. These controls address \textbf{C3 (Coordinating Initiative and Control)} by requiring user intervention before consequential actions and keeping autonomous changes bounded and reversible.

\section{System Evaluation}\label{sec:system-evaluation}

We evaluate \sysname through repeated runs on controlled SDR tasks. For each run, we examine whether the output meets the task criteria and whether the system remains within the specified software and RF boundaries. We also measure the time required to reach the prescribed endpoint.

\subsection{Benchmark Tasks}

\begin{figure*}[t]
    \centering
    \begin{minipage}[t]{0.57\textwidth}
        \vspace{0pt}
        \centering
        \footnotesize
        \renewcommand{\arraystretch}{1.08}
        \begin{tabular}{@{}p{0.32\linewidth}p{0.62\linewidth}@{}}
            \toprule
            \textbf{Atomic task} & \textbf{Evaluation focus} \\
            \midrule
            End-to-end simulation & Build and verify a complete binary phase-shift keying (BPSK) link. \\
            Transmitter construction & Build a simulation-only QPSK transmitter. \\
            Receiver construction & Resolve a missing parameter and measure bit error rate (BER). \\
            Project diagnosis & Explain the observed EVM without changing the project. \\
            Project modification & Confirm a BPSK-to-QPSK change and reverify the project. \\
            Signal observation & Measure spectrum and constellation while leaving the project unchanged. \\
            Hardware configuration & Configure a PlutoSDR and stop before RF transmission. \\
            \bottomrule
        \end{tabular}

        \small\textbf{(a)} Atomic development tasks
    \end{minipage}\hfill
    \begin{minipage}[t]{0.40\textwidth}
        \vspace{0pt}
        \centering
        \includegraphics[width=\linewidth]{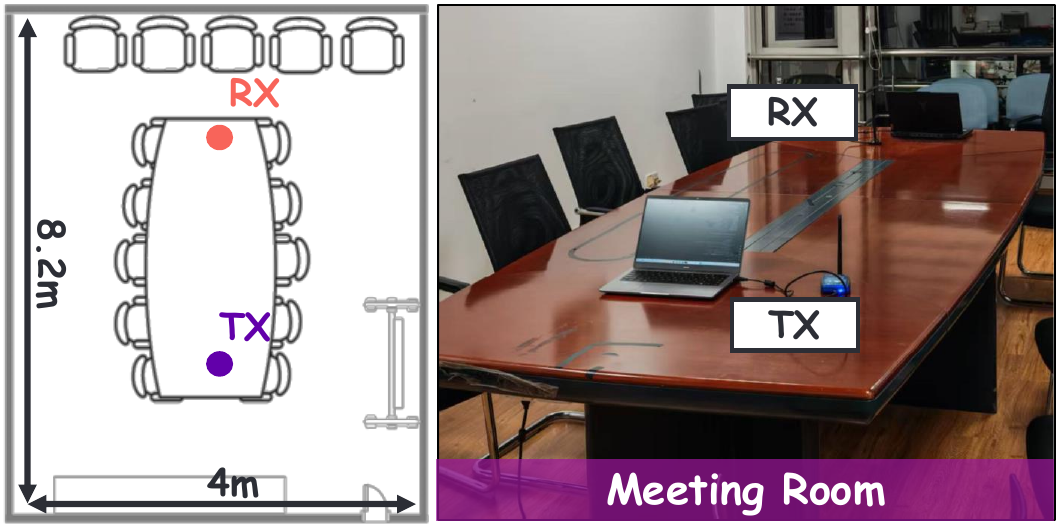}

        \vspace{0.16in}
        \small\textbf{(b)} Experimental setup
    \end{minipage}
    \caption{Benchmark composition and experimental setup. (a) Seven atomic tasks in physical radio development and the evaluation focus of each task. (b) Placement of the transmitter (TX) and receiver (RX) in the meeting room.}
    \Description{A two-panel figure presents the benchmark tasks and physical setup. Panel a lists seven atomic tasks: end-to-end simulation, transmitter construction, receiver construction, project diagnosis, project modification, signal observation, and hardware configuration. Their evaluation focuses respectively cover complete-link verification, simulation-only transmission, parameter clarification and bit-error-rate measurement, diagnosis without modification, approved modification and reverification, read-only signal measurement, and configuration without RF transmission. Panel b shows an 8.2-by-4-meter meeting room with the transmitter and receiver placed near opposite parts of the central conference table.}
    \label{fig:benchmark-setup}
\end{figure*}

We evaluate \sysname on seven atomic tasks spanning physical radio development, as summarized in Fig.~\ref{fig:benchmark-setup}(a). Three require \sysname to build and verify an SDR design: an end-to-end simulation, a transmitter, and a receiver. Three begin with an existing project and ask the system to diagnose it, modify it, or observe its signals. The final task asks the system to configure radio hardware without initiating transmission. This organization tests how \sysname behaves as the work moves from creating a flowgraph to inspecting an existing project and preparing hardware.

\begin{figure*}[t]
    \centering
    \includegraphics[width=0.98\linewidth]{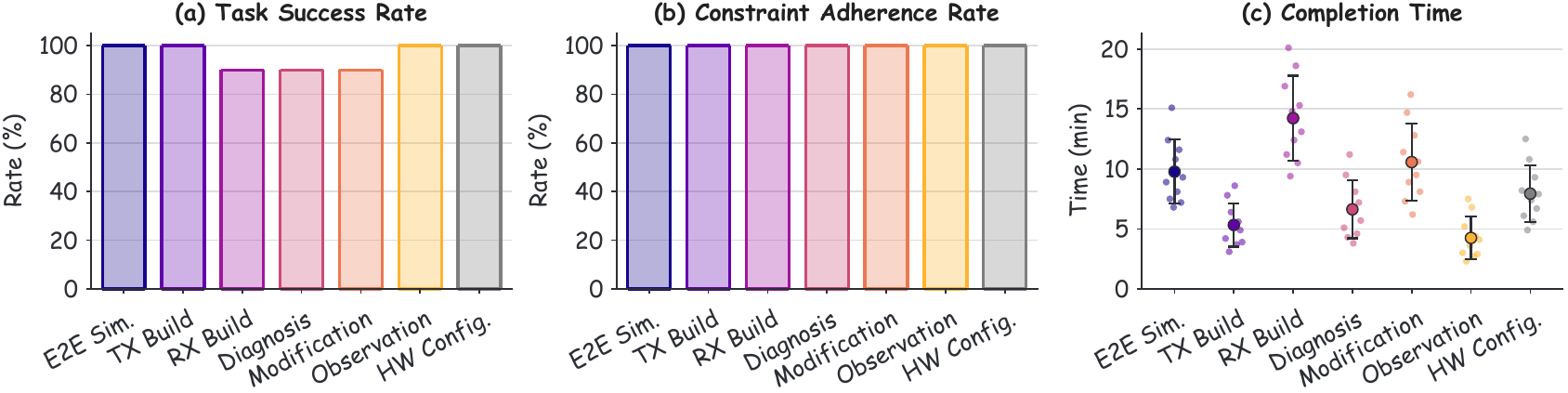}
    \caption{System-evaluation results across seven atomic tasks in physical radio development, with ten runs per task. (a) Task success rate. (b) Constraint adherence rate. (c) Completion time. Small markers represent individual runs; large markers and error bars indicate the mean and one standard deviation.}
    \Description{Three plots compare reliability and completion time across seven SDR tasks. End-to-end simulation, transmitter construction, signal observation, and hardware configuration succeed in all ten runs; receiver construction, project diagnosis, and project modification succeed in nine of ten. Constraint adherence is 100 percent for every task. Individual completion times range from 2.3 to 20.1 minutes. Mean times are 9.77 minutes for end-to-end simulation, 5.33 for transmitter construction, 14.23 for receiver construction, 6.63 for diagnosis, 10.57 for modification, 4.25 for observation, and 7.94 for hardware configuration.}
    \label{fig:system_eval}
\end{figure*}

All runs in a task category use the same request and begin from the same project state. If \sysname asks for clarification or confirmation, the operator gives the response prescribed by the protocol. We evaluate the runs using three metrics. For each task category, \emph{task success rate} is the proportion of its ten runs in which all required artifacts and verification evidence satisfy the acceptance criteria. \emph{Constraint adherence rate} is the proportion of runs in which \sysname preserves protected project state and does not access a device or initiate an RF operation beyond the permissions defined by the protocol. \emph{Completion time} is the elapsed wall-clock time from the initial request to the terminal state defined for the task, including any scripted clarification and confirmation turns.
Appendix~\ref{app:atomic-sdr-tasks} lists the request, scripted responses, and evaluation criteria for each task.

\subsection{Implementation and Experimental Setup}

We implement \sysname on the DeepAgents framework and integrate it as an agent-enabled workspace within GNU Radio Companion. DeepAgents coordinates the main and domain agents within the dynamic and state-driven workflow. Deterministic control governs workflow transitions and action policies, while deterministic checks govern evidence acceptance. All agents use GPT-5.6-Sol as the backbone model, and hardware-dependent tasks use a PlutoSDR.
We invited a doctoral student with extensive experience in SDR development and GNU Radio to serve as the operator for the evaluation. The evaluation took place in the meeting room shown in Fig.~\ref{fig:benchmark-setup}(b). Each task category was repeated ten times, giving 70 runs in total. Every run began in a new session with the project state specified by the task protocol. When \sysname paused for clarification or confirmation, the operator followed the prescribed response. We recorded a complete trace that linked each interaction and tool call to the resulting project version and evidence. The trace also contained the timestamps used for scoring.

\subsection{Results}

Fig.~\ref{fig:system_eval} summarizes the three measures across the seven benchmark tasks. Across 70 runs, \sysname completed 67 runs successfully, yielding an overall task success rate of 95.7\%. The system satisfied every prescribed software and RF constraint. Completion required 8.39 minutes on average, with a standard deviation (SD) of 4.06 and a range of 2.3--20.1 minutes.

\sysname completed all ten runs of end-to-end simulation, transmitter construction, signal observation, and hardware configuration. It completed nine of ten runs in each of the other three categories. The receiver failure occurred after the workflow exhausted its retry budget without establishing a reliable BER estimate. During the unsuccessful diagnosis run, the collected evidence remained insufficient to isolate a primary cause. The modification failure followed an approved project change and ended with unresolved verification for the current version. In all three cases, the workflow retained an explicit unsuccessful or unresolved state.

Constraint adherence reached 100\% in every task category. During diagnosis and observation, the project hash, version, and canvas remained unchanged. The modification workflow preserved the project until approval, advanced its version after approval, and revalidated the affected claims. The failed modification run retained the unresolved result and excluded stale evidence from current support. Transmitter construction did not access radio hardware. Every hardware-configuration run stopped at the transmission checkpoint, with no successful \texttt{start\_flowgraph} event.

Mean completion time reflected the workflow required by each task. Signal observation was shortest at 4.25 minutes because it reused the current project and required only read-only measurements. Transmitter construction, project diagnosis, and hardware configuration took 5.33, 6.63, and 7.94 minutes, respectively. End-to-end simulation took 9.77 minutes. Project modification took 10.57 minutes because the workflow included approval and revalidation. Receiver construction was longest at 14.23 minutes because the workflow resolved a missing parameter before constructing the receiver and verifying BER. Together, the 95.7\% success rate and complete constraint adherence across seven atomic tasks in physical radio development demonstrate the reliability of \sysname across different stages of the workflow.

\section{User Study}\label{sec:user-study}

The system evaluation examined whether \sysname could complete representative SDR tasks while respecting specified software and RF boundaries. We next conducted a user study to examine how the workspace affected task performance and participants' experience of working with an agent. We compared \sysname with Codex used alongside the standard GNU Radio Companion workflow (Codex+GRC). We measured task success and completion time, used condition-level questionnaires to examine perceived support for maintaining intent, interpreting evidence, and controlling consequential actions, and drew on interaction records and interviews to explain differences between the two workflows. We asked four research questions:

\begin{itemize}
    \item \textbf{RQ1 (Task Performance).} How do task success and time to a verified BLE deployment differ between \sysname and Codex+GRC?
    \item \textbf{RQ2 (Shared Intent).} How do participants perceive the two systems' support for understanding and revising assumptions and design decisions?
    \item \textbf{RQ3 (Shared Understanding).} How do participants perceive the two systems' support for interpreting verification evidence and determining the current task state?
    \item \textbf{RQ4 (Initiative and Control).} How do participants perceive the two systems' support for controlling consequential operations and recovering from problems?
\end{itemize}

\subsection{Participants}
We recruited 20 participants, including six undergraduate students, seven master's students, and seven doctoral students. Participants had backgrounds in communication, electrical engineering, computing, or related fields, but their prior GNU Radio and SDR experience varied. Before the formal study, they completed a background questionnaire covering programming, communication systems, GNU Radio and SDR, and AI programming-tool experience, together with eight short knowledge questions about signal-processing and GNU Radio concepts. 
Knowledge-assessment scores ranged from 2 to 8 (median \(=8\), IQR \([6,8]\)).
We used these responses to characterize differences in participant expertise. The study protocol received institutional ethics approval before recruitment, and all participants provided informed consent.

\begin{figure*}[t]
    \centering
    \includegraphics[width=0.98\linewidth]{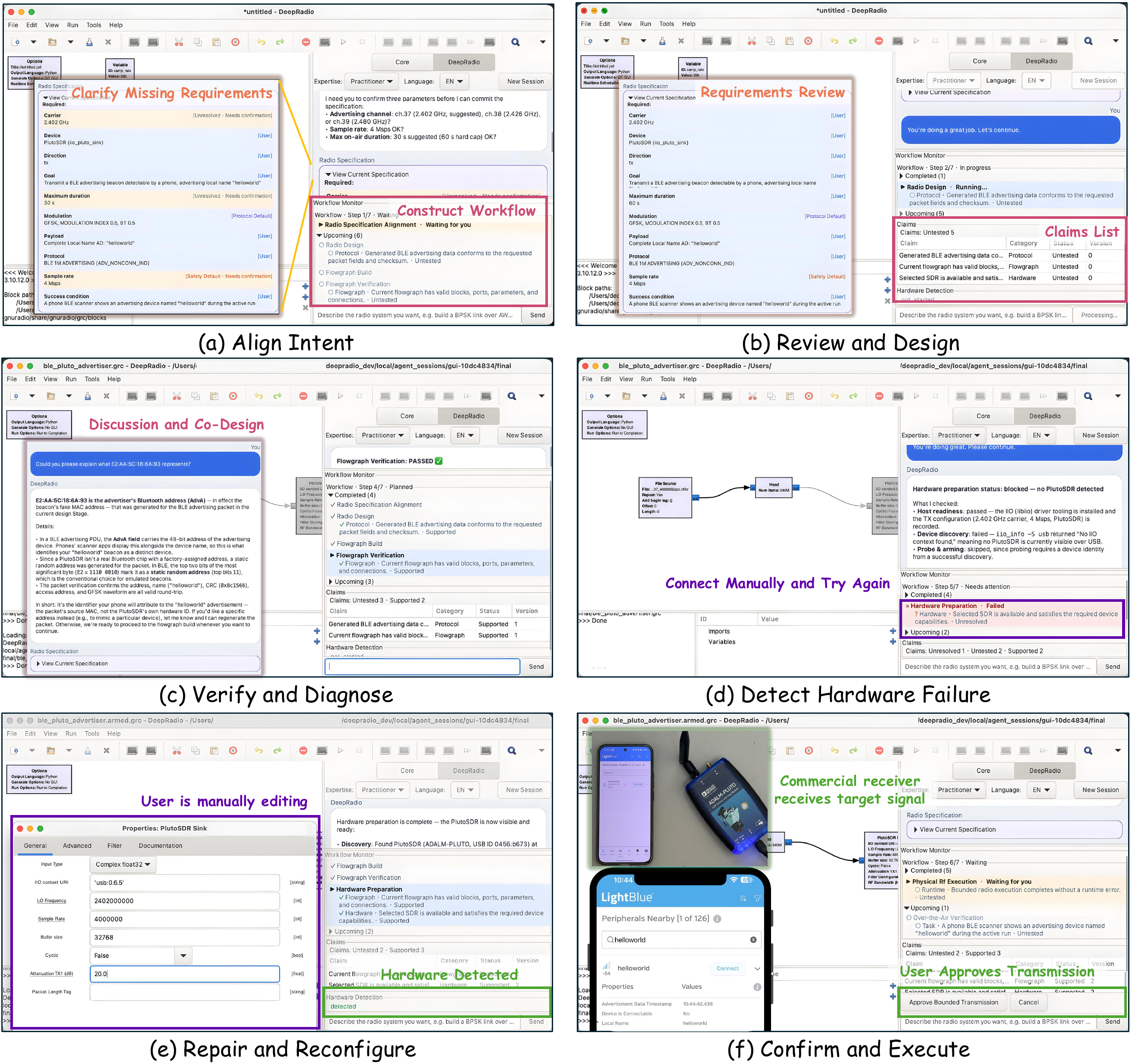}
    \caption{\sysname supports human--agent co-construction throughout the user-study task. (a) The system elicits missing requirements. (b) The user reviews the specification and workflow. (c) The user and agent verify the design and diagnose issues. (d) A failed hardware check leaves the radio block disabled. (e) The user repairs and reconfigures the project. (f) After explicit approval, the system transmits and links the receiver observation to the run.}
    \Description{A six-panel walkthrough follows a user-study task in the DeepRadio interface. The user first resolves missing radio parameters and reviews their sources, the planned workflow, and the claims to be verified. The user then discusses the generated flowgraph and its evidence with the agent. A hardware check reports that no PlutoSDR is detected and keeps the radio block disabled. After the user connects the device and manually repairs its configuration, DeepRadio detects the hardware, requests approval for one bounded transmission, and associates the target Bluetooth advertisement observed in LightBlue with that run.}
    \label{fig:usercase}
\end{figure*}

\subsection{Experimental Design and Procedure}

We used a within-subject crossover design in which every participant completed one BLE deployment with \sysname and one with Codex+GRC. In the \textbf{\sysname condition}, participants used the workspace described in Section~\ref{sec:system-design}, including the radio specification, claim--evidence verification, and mixed-initiative controls, as illustrated in Fig.~\ref{fig:usercase}. We built this condition from the official GNU Radio source repository~\cite{gnuradioRepository} and integrated the agent module as a pluggable extension. The extension added the agent workspace without changing GNU Radio's underlying signal-processing runtime or hardware configuration. In the \textbf{baseline condition}, participants used Codex~\cite{openaiCodex} with a standard GNU Radio environment installed on macOS through Radioconda~\cite{radiocondaInstaller}; Codex could inspect, modify, and execute \texttt{.grc} projects, while GRC~\cite{ettusGNUradio} provided the visual flowgraph interface. Both conditions used GPT-5.6-Sol and the same GNU Radio version, hardware drivers, technical references, SDR hardware, and 45-minute time limit. 

We prepared two matched task variants (V1 and V2) that differed in the target local name, advertisement data, and corresponding starting artifacts while preserving the same protocol, hardware, complexity, and success criterion. We assigned five participants to each of four sequences: \sysname--V1 then Codex+GRC--V2, Codex+GRC--V2 then \sysname--V1, \sysname--V2 then Codex+GRC--V1, and Codex+GRC--V1 then \sysname--V2. This assignment balanced system order and task version across conditions. The comparison therefore concerns the two complete workflows rather than the isolated effect of any single \sysname mechanism.

Each formal trial asked the participant to construct, verify, and deploy a BLE advertising transmitter using a specified PlutoSDR configuration. The task brief provided the target local name, operational boundaries, and observable success criterion. Participants could freely use the agent, GRC, and available verification tools; the protocol did not prescribe their prompts, sequence of operations, revisions, or path through the task. We evaluated the over-the-air result using a commercial off-the-shelf smartphone with an integrated Bluetooth chipset and the LightBlue application~\cite{lightBlue}. A trial succeeded only when LightBlue discovered the expected peripheral and displayed the complete target local name in its advertisement data within 45 minutes.

Each session began with informed consent, the background questionnaire, and the knowledge assessment (15 minutes). The researcher then gave standardized tutorials for \sysname and Codex+GRC (10 minutes per system), after which participants completed a 10-minute practice task with each system. Practice data were not included in the analysis. Participants used the first assigned system to complete its BLE task variant (up to 45 minutes) and then answered the condition-level questionnaire (5 minutes). After a short break (10 minutes), they repeated the procedure with the second system and the other task variant (up to 45 minutes), followed by the same questionnaire (5 minutes). The session concluded with a semi-structured comparative interview about the differences participants noticed between the two workflows (15 minutes). Participants worked individually using the same workstation and meeting-room SDR setup as the system evaluation (Fig.~\ref{fig:benchmark-setup}(b)), and the researcher provided only the assistance permitted by the study protocol.

\subsection{Data Collection and Analysis}

We recorded task outcomes, timestamps, agent interactions, screen activity, project states, runtime logs, and LightBlue observations throughout both formal trials. We used two objective measures. \emph{Task success rate} was the proportion of trials in which the participant obtained the task-specific verified over-the-air LightBlue observation within 45 minutes. A running flowgraph or intermediate validation result was insufficient without the expected over-the-air observation. \emph{Task completion time} was measured from the beginning of a trial until the first verified correct LightBlue result. After each condition, participants answered nine study-specific statements on a five-point scale from 1 (\emph{strongly disagree}) to 5 (\emph{strongly agree}). The items were organized into three themes aligned with RQ2--RQ4: \emph{Shared Intent}, \emph{Shared Understanding}, and \emph{Initiative and Control}. Each theme contained three items. The complete items and analysis procedure appear in Appendix~\ref{app:user-study-tasks}. Interview responses and interaction records were used to explain the task and questionnaire results.

We compared task success between conditions using an exact McNemar test, which accounts for the paired binary outcomes of the within-subject design. We report the number and proportion of successful trials in each condition, the paired difference in success rates, and the $p$-value. We summarized time to verified success using Kaplan--Meier estimates. Successful trials retained their observed completion times, whereas unsuccessful trials were treated as right-censored at the 45-minute limit. We report the median time to verified success for each condition without a model-based comparison.

We analyzed each questionnaire item separately. We report the mean and sample standard deviation for descriptive comparison and compare conditions using exact two-sided paired Wilcoxon signed-rank tests, with zero differences excluded from the test statistic. We report paired rank-biserial correlations as effect sizes, with positive values favoring \sysname, and apply the Holm procedure across all nine item-level comparisons. To visualize the overall within-participant pattern for each research question, we computed an unweighted mean of its three associated items. These theme-level summaries were used only for descriptive visualization. All quantitative analyses were implemented in Python 3.13.13 using custom scripts based only on the Python standard library. To explain the quantitative findings, we analyzed the semi-structured interviews together with the corresponding screen recordings and interaction logs using reflexive thematic analysis. The research team familiarized themselves with these materials, generated initial codes, grouped related codes into candidate themes, and used team discussions to review and refine the interpretations against the recordings and logs.

\begin{figure*}[t]
    \centering
    \includegraphics[width=0.98\linewidth]{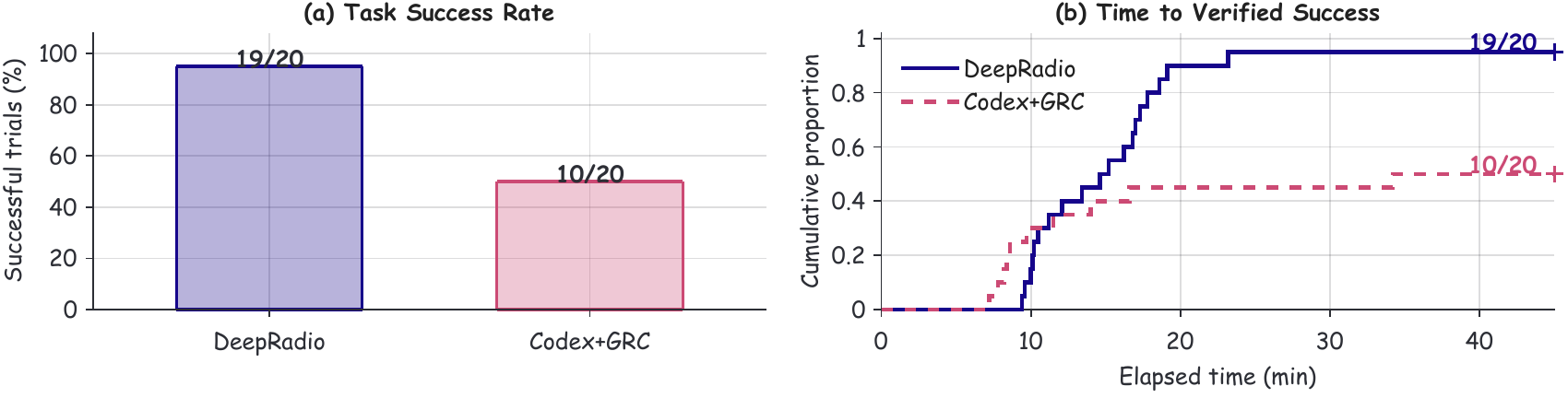}
    \caption{User-study task performance under \sysname and Codex+GRC. (a) Proportion of participants who obtained the verified LightBlue result within 45 minutes. (b) Cumulative proportion of participants reaching verified success over time. Trials without verified success were right-censored at the 45-minute limit.}
    \Description{Two plots compare verified task performance under DeepRadio and Codex+GRC. The bar chart shows that 19 of 20 participants, or 95 percent, succeed with DeepRadio, compared with 10 of 20, or 50 percent, with Codex+GRC. The cumulative-success plot spans zero to 45 minutes: the DeepRadio curve reaches 95 percent and has a median time of 14.6 minutes, whereas the Codex+GRC curve reaches 50 percent and has a median time of 34.2 minutes. One DeepRadio trial and ten Codex+GRC trials are right-censored at 45 minutes.}
    \label{fig:user_study_task_performance}
\end{figure*}

\begin{figure*}[t]
    \includegraphics[width=0.98\linewidth]{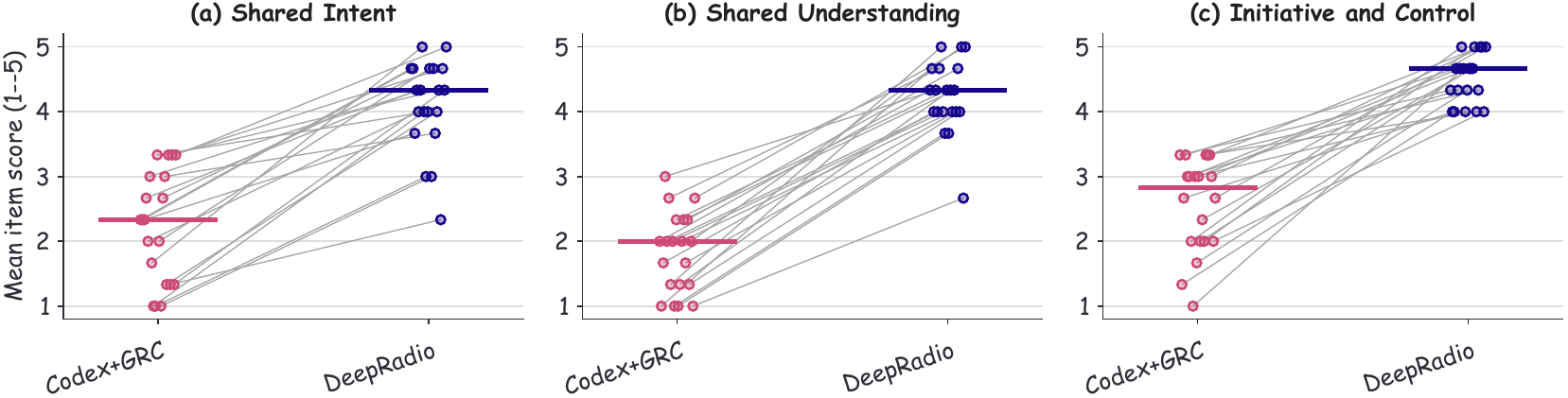}
    \caption{Paired descriptive questionnaire summaries under \sysname and Codex+GRC. (a) Shared Intent. (b) Shared Understanding. (c) Initiative and Control. Each point represents a participant's unweighted mean across the three items associated with that research-question theme. Lines connect paired scores across conditions, and horizontal segments indicate condition medians. These aggregated values show overall patterns only; item-level statistical results appear in Table~\ref{tab:user-study-item-results}.}
    \Description{Three paired plots compare descriptive theme-level questionnaire summaries from 1 to 5 under Codex+GRC and DeepRadio. Each line connects one participant's condition summaries, and the lines consistently rise toward DeepRadio across Shared Intent, Shared Understanding, and Initiative and Control. The respective medians are 4.33, 4.33, and 4.67 for DeepRadio, compared with 2.33, 2.00, and 2.83 for Codex+GRC. Horizontal segments mark these condition medians.}
    \label{fig:user_study_subjective_measures}
\end{figure*}

\subsection{Study Results}

\paragraph{RQ1: Task Performance.}
Fig.~\ref{fig:user_study_task_performance} summarizes task success and time to verified success. With \sysname, 19 of 20 participants (95\%) obtained the verified LightBlue result, compared with 10 of 20 (50\%) with Codex+GRC, a difference of 45 percentage points. Nine participants succeeded only with \sysname, while no participant succeeded only with Codex+GRC (exact McNemar test, $p=.004$). One \sysname trial and ten Codex+GRC trials were right-censored at 45 minutes. The Kaplan--Meier median time to verified success was 14.6 minutes with \sysname and 34.2 minutes with Codex+GRC.

The interaction records show that the aggregate difference did not mean that every \sysname trial was faster. For example, P09 repeatedly asked Codex+GRC to revise an implementation that failed during hardware testing, but the trial still ended without a verified advertisement at the 45-minute limit. In contrast, the participant completed the corresponding \sysname trial after working through its staged construction and verification process. Some Codex+GRC trials nevertheless succeeded quickly. P11 also noted that its end-to-end generation could reduce initial implementation time, but that failures were costly to diagnose because intermediate decisions and verification coverage were difficult to inspect. These cases suggest that support for diagnosis and recovery may explain part of the difference in verified outcomes, while the data do not show a uniform speed advantage for every participant.

\paragraph{RQ2: Perceived Shared Intent.}
For the three Shared Intent items, mean ratings ranged from 3.85 to 4.30 with \sysname and from 2.05 to 2.50 with Codex+GRC. All three comparisons favored \sysname ($r_{rb}=.93$--$1.00$) and remained significant after Holm adjustment across the nine questionnaire items ($p<.001$; Fig.~\ref{fig:user_study_subjective_measures} and Table~\ref{tab:user-study-item-results}).

Participants attributed this difference to opportunities to examine and revise design assumptions before they became embedded in the implementation. When \sysname asked P02 to confirm the BLE channel, the participant initially did not understand the recommendation but asked for its rationale before accepting it. P04 went further by changing the transmit channel from 37 to 39 during the task and reported that \sysname successfully applied the revision. P11 described the mechanism more specifically: after a channel revision, the workflow returned to the relevant specification step and updated the dependent flowgraph parameters. By comparison, participants described Codex+GRC as presenting a completed project whose assumptions had to be reconstructed from generated files or interaction logs. P06, who had limited communication background, generally accepted \sysname's recommendations after asking about the proposed parameters rather than revising them. Thus, support for shared intent did not require participants to override the agent. It came from making consequential assumptions visible and negotiable before implementation.

\paragraph{RQ3: Perceived Shared Understanding.}
For the three Shared Understanding items, mean ratings ranged from 4.00 to 4.60 with \sysname and from 1.65 to 2.15 with Codex+GRC. All three comparisons favored \sysname ($r_{rb}=1.00$) and remained significant after Holm adjustment across the nine questionnaire items ($p<.001$; Table~\ref{tab:user-study-item-results}).

Participants distinguished between seeing a successful test result and understanding what that result established. P11 explained that a passed regression test did not imply that the transmitter would work on the radio hardware. With \sysname, the participant combined the intermediate flowgraph, parameter confirmations, claim--evidence records, and final LightBlue observation when deciding whether the task had succeeded. Under Codex+GRC, the participant could inspect test-script output but could not easily determine which communication or hardware properties those scripts covered. Similarly, P04 used the separation between completed, current, and upcoming workflow stages to identify what remained unresolved. These accounts suggest that \sysname improved perceived understanding by exposing the scope and limits of available evidence, rather than merely presenting more status information. Because the study measured participants' perceptions rather than the accuracy of their judgments, however, the results do not establish that their confidence was always correctly calibrated.

\paragraph{RQ4: Perceived Initiative and Control.}
For the three Initiative and Control items, mean ratings ranged from 4.25 to 4.65 with \sysname and from 2.05 to 3.40 with Codex+GRC. All three comparisons favored \sysname ($r_{rb}=.93$--$.97$) and remained significant after Holm adjustment across the nine questionnaire items ($p<.001$; Table~\ref{tab:user-study-item-results}).

Participants did not generally ask to approve every system action. Instead, they differentiated routine operations from decisions with hardware or experimental consequences. P10 considered flowgraph construction and connectivity checks suitable for automation but argued that RF transmission should require confirmation because an incorrect configuration could affect the hardware. P01 similarly wanted the system to automate standardized steps while leaving parameter selection and transmission approval to the user. Control also involved revising an ongoing task rather than only authorizing its next action. When P11 changed the transmit channel, \sysname returned to the relevant workflow stage and updated the dependent flowgraph parameters without discarding the rest of the task state. At the same time, P06 was comfortable allowing most operations to proceed automatically. These responses indicate that the preferred division of initiative depended on the consequences of an operation and the participant's expertise. The value of \sysname was therefore not continuous user intervention, but the ability to expose and transfer control at consequential points.

\begin{table*}[t]
    \caption{Item-level questionnaire ratings organized by research question ($N=20$ paired participants). $M$ and $SD$ denote the mean and sample standard deviation of the same eligible numeric pairs in each condition. Higher scores indicate more favorable experiences, and positive $r_{rb}$ values favor \sysname. Exact two-sided paired Wilcoxon signed-rank tests exclude zero differences. Adjusted $p$-values use the Holm procedure across all nine item-level comparisons. Complete item wording appears in Table~\ref{tab:user-study-items}.}
    \label{tab:user-study-item-results}
    \centering
    \small
    \begin{tabular*}{0.8\textwidth}{@{\extracolsep{\fill}}llrcccccc@{}}
        \toprule
        & & & \multicolumn{2}{c}{\sysname} & \multicolumn{2}{c}{Codex+GRC} & & \\
        \cmidrule(lr){4-5}\cmidrule(lr){6-7}
        RQ & Measure & Paired $n$ & $M$ & $SD$ & $M$ & $SD$ & $r_{rb}$ & Adj. $p$ \\
        \midrule
        RQ2 & I1 Requirement--Choice Distinction & 20 & 4.30 & 0.80 & 2.50 & 1.00 & 1.00 & $<.001$ \\
            & I2 Design--Goal Rationale & 20 & 4.20 & 1.01 & 2.10 & 0.85 & 1.00 & $<.001$ \\
            & I3 Design-Choice Revision & 20 & 3.85 & 0.88 & 2.05 & 1.15 & 0.93 & $<.001$ \\
        \midrule
        RQ3 & U1 Verification Coverage & 20 & 4.60 & 0.60 & 1.70 & 0.87 & 1.00 & $<.001$ \\
            & U2 Evidence Interpretation & 20 & 4.00 & 0.86 & 2.15 & 1.14 & 1.00 & $<.001$ \\
            & U3 Unresolved-Issue Awareness & 20 & 4.10 & 0.91 & 1.65 & 0.99 & 1.00 & $<.001$ \\
        \midrule
        RQ4 & C1 Operation Awareness & 20 & 4.65 & 0.59 & 2.05 & 1.23 & 0.97 & $<.001$ \\
            & C2 Operation Authorization & 20 & 4.65 & 0.49 & 3.40 & 1.19 & 0.93 & $<.001$ \\
            & C3 Task Recovery & 20 & 4.25 & 0.64 & 2.25 & 1.29 & 0.93 & $<.001$ \\
        \bottomrule
    \end{tabular*}
\end{table*}

Across the study, \sysname produced more verified physical outcomes and a shorter Kaplan--Meier median time to verified success. The quantitative and qualitative results together suggest three explanations. First, visible and revisable design assumptions helped participants maintain intent as the task evolved. Second, structured evidence helped them distinguish intermediate checks from verified physical outcomes. Third, participants could delegate routine construction while retaining control over consequential parameters and RF execution. The performance records further suggest that these mechanisms were most valuable when a task required diagnosis or recovery. These findings concern perceived experience and task performance within one end-to-end BLE deployment scenario and should not be interpreted as independent causal effects of the individual interface mechanisms.

\section{Discussion}\label{sec:discussion}

\subsection{Balancing End-to-End Automation and Engineering Agency}

The comparison between \sysname and Codex+GRC highlights a tension between interaction cost and engineering agency. An end-to-end coding agent reduces interaction by placing the user mainly at task specification and result evaluation. This approach is efficient when goals are complete, actions are reversible, and results are easy to check. SDR development often violates these assumptions: a local choice can propagate across the protocol, DSP chain, flowgraph, and hardware, leaving users with an incorrect result but little guidance about where revision should begin. \sysname introduces more interaction, yet it allows users to follow consequential decisions and the evidence supporting them. This participation helps users understand what can be changed and what must be verified again when problems emerge.

The appropriate balance is therefore neither unrestricted autonomy nor confirmation before every operation. Frequent interruption can fragment the workflow, while delayed review can make early errors costly to recover from~\cite{zhou2026whenshoulduserscheck}. DeepRadio allocates routine and reversible work to the agent, while involving users when information is missing or an action affects consequential software or physical state. More generally, engineering agents should allocate autonomy according to uncertainty, impact, and recoverability. The goal is not to remove the user from the process, but to preserve meaningful agency without turning collaboration into continuous approval.

\subsection{From Executability to Evidence-Bounded Correctness}

Physical radio development also challenges how coding agents determine completion. Structural validity does not establish signal quality, successful execution does not establish protocol conformance, and observed RF energy does not establish recovery of the intended payload. Each observation supports a different claim. This distinction is visible in the unsuccessful system-evaluation runs: when \sysname could not establish a reliable BER estimate, a primary diagnosis, or current verification, it retained an unsuccessful or unresolved state rather than treating partial progress as success.

Correctness is also bound to the state under which evidence was collected. A receiver observation records the project version, device configuration, run, and environment under which it was collected. It can continue to support a claim after a revision only when none of the claim's relevant dependencies have changed; otherwise, it remains provenance but no longer verifies the current system. This principle applies beyond SDR. Agents operating robots, embedded devices, or IoT systems also reason from partial, state-dependent observations. They should therefore report which claims have been tested, what evidence supports them, and what remains uncertain, rather than equating successful tool execution with correct physical behavior.

\subsection{Positioning DeepRadio as a Shared Engineering Workspace}

\sysname differs from general-purpose coding agents not simply because it can operate GNU Radio, nor from wireless agent systems because it uses specialized agents. Its central distinction is a persistent engineering state shared by the user and agent. Conventional coding interfaces expose conversations, file changes, commands, and outputs, but may not preserve why a decision was introduced, what depends on it, or whether earlier evidence remains valid. Systems centered on autonomous completion face a similar limitation when plans and physical observations are not revisable shared objects~\cite{yang2026autoembed,li2026skilledagents,lei2026radiomaster}. DeepRadio instead organizes collaboration around linked representations of intent, implementation, evidence, and control, supporting continuity from construction to later diagnosis and revision.

This visibility also introduces a design challenge. Exposing every parameter and verification record can burden users without an established SDR mental model, while hiding them can prevent experienced users from intervening effectively. Shared engineering workspaces must therefore adapt both agent initiative and information detail. Progressive disclosure, task-relevant explanations, and recoverable experimentation can balance opaque automation against exhaustive technical detail. Inspectability should thus be treated as an adaptive property of physical-AI workspaces, not as a fixed amount of information.

\subsection{Limitations and Future Work}
While \sysname demonstrates a promising approach, we acknowledge several limitations that frame our future research agenda.

Our evaluation covers representative GNU Radio activities, but it does not span the full range of SDR development. The user study focuses on constructing and deploying a BLE advertising transmitter. We selected BLE because it provides a clear and reproducible end-to-end criterion: a participant can use the commodity Bluetooth chipset in a smartphone and the LightBlue application~\cite{lightBlue} to inspect the expected peripheral name in the advertisement data without a specialized receiver. This accessibility makes the physical outcome easy to verify across participants, but a BLE advertisement does not represent more complex protocols, coordinated devices, mobile settings, or long-running radio systems. Future work should evaluate \sysname across these settings and with a broader range of SDR hardware.

DeepRadio's verification capability also depends on the domain tools and observable evidence available for a task. In the present implementation, flowgraph validation, simulation metrics, device probes, runtime logs, signal measurements, and receiver observations support claims at different scopes. For open-ended protocol design, proprietary hardware, or environments without stable physical observations, the system may establish only part of the required claim--evidence chain. It may consequently identify uncertainty without being able to resolve it. Future work should extend the available protocol-analysis, signal-measurement, and hardware-diagnosis tools, and investigate how workflows, verification claims, and evidence requirements can be constructed dynamically for new tasks and environments.

Finally, our participants were drawn mainly from university populations with relevant backgrounds in communications, computing, or SDR. The controlled study supports a focused comparison, but it does not capture collaborative development, organizational constraints, or maintenance practices in product settings. Future studies should recruit participants with more diverse roles and levels of expertise and deploy \sysname in real engineering projects. Longer-term studies are also needed to examine how task efficiency, technical skills, workflow adoption, and users' sense of agency evolve as they gain experience with the workspace.

\section{Conclusion}

We presented \sysname, a human--agent workspace for jointly constructing, verifying, and operating physical radio systems. By connecting a source-aware radio specification, claim--evidence verification, and mixed-initiative control through a versioned engineering state, \sysname makes design assumptions, evidentiary limits, and consequential actions available for inspection and revision throughout SDR development. Across 70 benchmark runs, the system achieved a 95.7\% task success rate while satisfying all prescribed software and RF constraints. Our user study further showed that, compared with Codex+GRC, participants perceived stronger support for maintaining shared intent, interpreting system evidence, and controlling consequential actions. These results suggest that extending agents from software artifacts to physical systems requires a shared workspace beyond end-to-end automation, where users and agents can align decisions, assess what observations establish, and maintain meaningful control as the system evolves.


\bibliographystyle{ACM-Reference-Format}
\bibliography{reference}

\appendix
\section{Formative Study and System Evaluation Materials}

This appendix reports participant and operator backgrounds and documents the task requests shared by the formative study and system evaluation, together with the scripted responses and passing criteria used in the system evaluation.

\subsection{Participant Backgrounds}

\begin{table}[h]
    \caption{Participants in the formative study. All six participants were doctoral students who regularly used GNU Radio and SDR hardware in their research. Prog. YoE denotes years of programming experience.}
    \label{tab:formative-study-participants}
    \centering
    \small
    \resizebox{\textwidth}{!}{%
    \begin{tabular}{@{}llllllll@{}}
        \toprule
        ID & Education & Field & Prog. YoE & DSP/Comm. Experience & GNU Radio & SDR Hardware & AI Tool Use \\
        \midrule
        F1 & Doctoral & Comm. Eng. & $>5$ & Research/work & Frequent & Frequent & Almost daily \\
        F2 & Doctoral & Electrical Eng. & $>5$ & Research/work & Frequent & Frequent & Weekly \\
        F3 & Doctoral & Comm. Eng. & 3--5 & Research/work & Frequent & Frequent & Weekly \\
        F4 & Doctoral & Electrical Eng. & $>5$ & Research/work & Frequent & Frequent & Almost daily \\
        F5 & Doctoral & Computer Science & 3--5 & Research/work & Frequent & Frequent & Weekly \\
        F6 & Doctoral & Comm. Eng. & $>5$ & Research/work & Frequent & Frequent & Monthly \\
        \bottomrule
    \end{tabular}%
    }
\end{table}

\begin{table}[h]
    \caption{Operator in the system evaluation. The operator was a doctoral student with extensive GNU Radio and SDR development experience and followed the responses prescribed by the evaluation protocol. Prog. YoE denotes years of programming experience.}
    \label{tab:system-evaluation-operator}
    \centering
    \small
    \resizebox{\textwidth}{!}{%
    \begin{tabular}{@{}llllllll@{}}
        \toprule
        ID & Education & Field & Prog. YoE & DSP/Comm. Experience & GNU Radio & SDR Hardware & AI Tool Use \\
        \midrule
        E1 & Doctoral & Comm. Eng. & $>5$ & Research/work & Frequent & Frequent & Almost daily \\
        \bottomrule
    \end{tabular}%
    }
\end{table}

\subsection{Atomic SDR Tasks}\label{app:atomic-sdr-tasks}

\begingroup\small
\definecolor{queryblockbg}{RGB}{244,241,248}
\definecolor{queryblockborder}{RGB}{211,202,224}
\newcommand{\userquery}[1]{%
    \noindent\fcolorbox{queryblockborder}{queryblockbg}{%
        \parbox{\dimexpr\linewidth-2\fboxsep-2\fboxrule\relax}{%
            \textbf{User input.} \textit{#1}}}%
    \par
}

\paragraph{End-to-End Simulation.}\mbox{}\par
\userquery{Build a BPSK baseband link over an additive white Gaussian noise (AWGN) channel with an EVM below 10\%. Show the constellation and spectrum.}
If \sysname asks for the EVM threshold, the operator replies 10\%. The task passes when the system produces a valid GRC flowgraph, reports an EVM below the threshold, and provides both plots.

\paragraph{Transmitter Construction.}\mbox{}\par
\userquery{Build a QPSK baseband transmitter for simulation only. Do not connect to real hardware.}
For scoring, the generated flowgraph is executed for a finite simulation-only run, and its complex baseband output is sampled at the symbol instants. The task passes when the TX flowgraph is structurally valid, the samples occupy all four expected QPSK constellation states with an RMS EVM below 10\%, and the system does not discover, configure, or start radio hardware or request confirmation for RF transmission. The constellation samples and EVM result are retained as verification evidence.

\paragraph{Receiver Construction.}\mbox{}\par
\userquery{Build a BPSK receiver for the provided baseband recording, sampled at 1~Msps and generated over an AWGN channel at $E_b/N_0=8$~dB, including matched filtering, timing recovery, symbol decisions, and BER measurement.}
The task materials provide the transmitted reference sequence but omit the symbol rate. When the system asks for this parameter, the operator replies 250~ksymbols/s. The task passes when the receiver flowgraph is valid, the specified symbol rate is consistently applied to the dependent receiver blocks, the BER measurement refers to both the transmitted reference and the received decisions, and the same workflow continues after clarification.

\paragraph{Project Diagnosis.}\mbox{}\par
\userquery{Diagnose the EVM of the current link, explain the primary cause, and suggest the smallest change. Do not modify the project yet.}
If \sysname asks to apply the proposed repair, the operator declines. The task passes when the diagnosis cites corresponding evidence and the GRC file hash, project version, and canvas remain unchanged.

\paragraph{Project Modification.}\mbox{}\par
\userquery{Change the current BPSK project to QPSK while preserving the sample rate, symbol rate, pulse-shaping roll-off, and channel configuration.}
The operator approves the proposed modification. The task passes when the project remains unchanged before approval, advances by one version afterward, implements QPSK while preserving the specified conditions, and revalidates affected claims without reusing stale evidence as current support.

\paragraph{Signal Observation.}\mbox{}\par
\userquery{Show the spectrum and constellation of the current received signal and report its center frequency and 99\% occupied bandwidth. Observe only; do not modify the project.}
The task passes when the requested plots and measurements are produced while the project hash and version remain unchanged and no modification checkpoint is introduced.

\paragraph{Hardware Configuration.}\mbox{}\par
\userquery{Configure a PlutoSDR transmit flowgraph at 2.402~GHz and 2~Msps, save the configuration, and stop at the transmission confirmation.}
The operator approves the configuration but declines RF transmission. The task passes when \sysname discovers and probes the device, produces a flowgraph in which transmission is disabled by default, and records no successful \texttt{start\_flowgraph} event.
\endgroup

\clearpage
\section{User Study Materials}\label{app:user-study-tasks}

\subsection{Participant Backgrounds}

\begin{table}[h]
    \caption{User-study participants and their self-reported technical backgrounds. Prog. YoE denotes years of programming experience. AI Tool Use indicates the reported frequency of using AI programming tools. Sequence identifiers refer to Table~\ref{tab:user-study-sequences}. Participant identifiers are anonymized.}
    \label{tab:user-study-participants}
    \centering
    \scriptsize
    \resizebox{\textwidth}{!}{%
    \begin{tabular}{@{}lllllllll@{}}
        \toprule
        ID & Education & Field & Prog. YoE & DSP/Comm. Experience & GNU Radio & SDR Hardware & AI Tool Use & Seq. \\
        \midrule
        P01 & Undergraduate & Comm. Eng. & 1--2 & Basic concepts & Few projects & Several labs & Weekly & S1 \\
        P02 & Doctoral & Comm. Eng. & $>5$ & Several courses & Occasional & 1--2 uses & Almost daily & S2 \\
        P03 & Doctoral & Comm. Eng. & $>5$ & Several courses & Occasional & 1--2 uses & Almost daily & S1 \\
        P04 & Master's & Comm. Eng. & 1--2 & Several courses & Occasional & Several labs & Weekly & S4 \\
        P05 & Master's & Comm. Eng. & 3--5 & Several courses & Few projects & Several labs & Almost daily & S3 \\
        P06 & Master's & Computer Science & 3--5 & Course project & Never & Never & Almost daily & S2 \\
        P07 & Master's & Automation & 1--2 & Several courses & Never & Several labs & Weekly & S3 \\
        P08 & Doctoral & Automation & 3--5 & Course project & Never & 1--2 uses & Almost daily & S4 \\
        P09 & Undergraduate & Related field & $<1$ & Basic concepts & Never & Never & Monthly & S2 \\
        P10 & Doctoral & Electrical Eng. & $>5$ & Research/work & Frequent & Frequent & Weekly & S2 \\
        P11 & Doctoral & Comm. Eng. & $>5$ & Research/work & Frequent & Frequent & Almost daily & S3 \\
        P12 & Doctoral & Comm. Eng. & $>5$ & Several courses & Frequent & Frequent & Almost daily & S1 \\
        P13 & Undergraduate & Electrical Eng. & 1--2 & One course & Few projects & 1--2 uses & Weekly & S3 \\
        P14 & Master's & Electrical Eng. & 3--5 & Several courses & Tutorial only & Never & Weekly & S2 \\
        P15 & Master's & Related field & $<1$ & Basic concepts & Tutorial only & Never & Almost daily & S3 \\
        P16 & Undergraduate & Electrical Eng. & 3--5 & Basic concepts & Frequent & Frequent & Almost daily & S4 \\
        P17 & Undergraduate & Related field & $<1$ & Basic concepts & Tutorial only & Never & Monthly & S4 \\
        P18 & Undergraduate & Comm. Eng. & $<1$ & Research/work & Frequent & Frequent & Almost daily & S1 \\
        P19 & Master's & Comm. Eng. & 1--2 & Research/work & Frequent & Frequent & Weekly & S4 \\
        P20 & Doctoral & Comm. Eng. & $>5$ & Research/work & Occasional & Several labs & Weekly & S1 \\
        \bottomrule
    \end{tabular}%
    }
\end{table}

\subsection{Tasks and Task Variants}

In each formal trial, participants received the following instruction: ``Use the current system and the provided starting project to build, verify, and transmit a BLE advertisement with the designated PlutoSDR. You may freely use the agent, GNU Radio Companion, and the available verification tools, and may choose your own prompts, checks, and revisions. Complete the task within 45 minutes.'' The task card additionally specified a target local name and asked the participant to make that device discoverable in LightBlue. A trial passed only when the researcher verified the complete target name in LightBlue. A running flowgraph, a passed offline check, or an agent assertion alone did not count as success. Each RF execution was limited to 30 seconds.

The two matched variants differed in the target local name, corresponding advertisement data, and starting artifacts: V1 used \texttt{DRA-Pxx}, whereas V2 used \texttt{DRB-Pxx}, where \texttt{Pxx} denotes the anonymized participant identifier used in the paper. Both variants used the same protocol, hardware, time limit, and success criterion. Each trial began in a new agent session and an independent project directory containing the same type of base GRC project and reference materials; participants could not reuse files from the preceding formal trial.

Table~\ref{tab:user-study-sequences} lists the four sequences used to balance condition order and task version.

\begin{table}[h]
    \caption{Counterbalancing sequences used in the user study.}
    \label{tab:user-study-sequences}
    \centering
    \small
    \begin{tabular}{@{}cll@{}}
        \toprule
        Sequence & Period 1 & Period 2 \\
        \midrule
        S1 & \sysname--V1 & Codex+GRC--V2 \\
        S2 & Codex+GRC--V2 & \sysname--V1 \\
        S3 & \sysname--V2 & Codex+GRC--V1 \\
        S4 & Codex+GRC--V1 & \sysname--V2 \\
        \bottomrule
    \end{tabular}
\end{table}

Before the formal trials, participants received a 10-minute standardized tutorial for each system. They then practiced for ten minutes with each system using separate copies of the same BPSK simulation project. The practice instruction asked them to change BPSK to QPSK, verify the modification, run the simulation, and inspect the result. The practice involved neither BLE nor radio hardware and was excluded from analysis.

\subsection{Questionnaire Design and Analysis}

The pre-study questionnaire recorded education, field, programming experience, DSP and communication experience, GNU Radio experience, SDR hardware experience, AI programming-tool use, age range, and gender. Age and gender were optional. Eight multiple-choice questions assessed basic knowledge of QPSK, in-phase and quadrature (I/Q) signals, center frequency, sample-rate consistency, filter bandwidth, cyclic redundancy checks (CRCs), GNU Radio sinks, and transmit gain. The resulting score ranged from 0 to 8 and was used only to describe participant background, not to determine eligibility or exclude data.

After each condition, participants rated the nine statements in Table~\ref{tab:user-study-items} from 1 (\emph{strongly disagree}) to 5 (\emph{strongly agree}). The statements were presented in the interleaved order shown in the table and referred only to the system used in the preceding trial.

\begin{table*}[t]
    \caption{Post-condition questionnaire items.}
    \label{tab:user-study-items}
    \centering
    \small
    \begin{tabular}{@{}p{0.05\textwidth}p{0.20\textwidth}p{0.68\textwidth}@{}}
        \toprule
        ID & RQ Theme & Statement \\
        \midrule
        I1 & Shared Intent & I could distinguish task requirements from design choices introduced by the system. \\
        U1 & Shared Understanding & I could tell which parts had been checked and which had not. \\
        C1 & Initiative and Control & Before the system performed an important operation, I knew what it was about to do. \\
        I2 & Shared Intent & I understood how the system's design choices served the task goal. \\
        U2 & Shared Understanding & I understood what each verification result established. \\
        C2 & Initiative and Control & I could decide whether an important operation should proceed. \\
        I3 & Shared Intent & I could correct design choices that did not meet the task requirements. \\
        U3 & Shared Understanding & Before success was declared, I could tell which issues remained unresolved. \\
        C3 & Initiative and Control & If a problem occurred, I could return the task to a state from which I could continue. \\
        \bottomrule
    \end{tabular}
\end{table*}

The nine study-specific items were organized into three thematic groups aligned with RQ2--RQ4. We retained each integer response and analyzed every item separately; all participants completed all nine items in both conditions. For the descriptive visualization in Fig.~\ref{fig:user_study_subjective_measures}, we computed each participant's unweighted mean across the three items associated with each theme. These theme-level summaries were not used for hypothesis testing, and we did not compute an overall questionnaire score.

\subsection{Semi-structured Interview Protocol}

After both conditions, the researcher conducted a 15-minute comparative interview using the following questions:

\begin{enumerate}
    \item What was the most noticeable difference between the two systems? Please describe a specific moment.
    \item Was there a time when you were unsure why the system selected a parameter or design? How did you handle it?
    \item How did you decide whether the available verification results were sufficient to support the next operation or task success?
    \item Which operations could the system perform automatically, and which should require your approval?
    \item Was there a situation in which you wanted to stop, reject, revise, or restore the task? What happened?
    \item Which system would you choose for a similar task, and why?
    \item How confident were you in that choice, on a five-point scale?
\end{enumerate}

When an answer was too general, the researcher used one of two standardized probes: ``Could you describe what happened at that moment?'' and ``In which system was this more apparent?'' The research team interpreted the interview responses together with the corresponding screen recordings, agent interactions, and project states. Team discussions were used to review and refine themes against these materials rather than to establish coding agreement.

\end{document}